\documentclass[aps,showpacs,prl,twocolumn,superscriptaddress,apsrev,english]{revtex4-1}
\usepackage{multirow}
\usepackage{color}
\usepackage{amsfonts}
\usepackage{amsmath}
\usepackage{mathrsfs}
\usepackage{amssymb}
\usepackage{wasysym}
\usepackage{graphicx}
\usepackage{babel}
\usepackage{graphicx}
\usepackage{paralist}
\usepackage{indentfirst}
\usepackage{subfigure}
\usepackage{hyperref}
\usepackage{ulem}
\usepackage{soul}
\usepackage{cancel}
\usepackage{CJK}

\begin{document}
\begin{CJK*}{GBK}{}
\title{Rabi Spectroscopy and Sensitivity of a Floquet Engineered Optical Lattice Clock}
\author{Mo-Juan Yin}
\thanks{These authors contributed equally to this work.}
\affiliation{Key Laboratory of Time and Frequency Primary Standards, National Time Service Center, Chinese Academy of Sciences, Xi'an 710600, China}
\author{Tao Wang}
\thanks{These authors contributed equally to this work.}
\affiliation{Department of Physics, and Center of Quantum Materials and Devices, Chongqing University, Chongqing, 401331, China}
\author{Xiao-Tong Lu}
\affiliation{Key Laboratory of Time and Frequency Primary Standards, National Time Service Center, Chinese Academy of Sciences, Xi'an 710600, China}
\author{Ting Li}
\affiliation{Key Laboratory of Time and Frequency Primary Standards, National Time Service Center, Chinese Academy of Sciences, Xi'an 710600, China}
\author{Ye-Bing Wang}
\affiliation{Key Laboratory of Time and Frequency Primary Standards, National Time Service Center, Chinese Academy of Sciences, Xi'an 710600, China}
\author{Xue-Feng Zhang}
\email{zhangxf@cqu.edu.cn}
\affiliation{Department of Physics, and Center of Quantum Materials and Devices, Chongqing University, Chongqing, 401331, China}
\author{Wei-Dong Li}
\email{wdli@sxu.edu.cn}
\affiliation{Department of Physics and Institute of Theoretical Physics, Shanxi University, Taiyuan 030006, China}
\author{Augusto Smerzi}
\email{augusto.smerzi@ino.it}
\affiliation{Department of Physics and Institute of Theoretical Physics, Shanxi University, Taiyuan 030006, China}
\affiliation{QSTAR, INO-CNR, and LENS, Largo Enrico Fermi 2, I-50125 Firenze, Italy}
\author{Hong Chang}
\email{changhong@ntsc.ac.cn}
\affiliation{Key Laboratory of Time and Frequency Primary Standards, National Time Service Center, Chinese Academy of Sciences, Xi'an 710600, China}
\affiliation{School of Astronomy and Space Science, University of Chinese Academy of Sciences, Beijing 100049, China}
\begin{abstract}
We periodically modulate the lattice trapping potential of a $^{87}$Sr optical clock to Floquet engineer the clock transition. In the context of atomic gases in lattices, Floquet engineering has been used to shape the dispersion and topology of Bloch quasi-energy bands.
Differently from these previous works manipulating the external (spatial) quasi-energies, we target the internal atomic degrees of freedom. We shape Floquet spin quasi-energies and measure their resonance profiles with Rabi spectroscopy. We provide the spectroscopic sensitivity of each band by measuring the Fisher information and show that this is not depleted by the Floquet dynamical modulation. The demonstration that the internal degrees of freedom can be selectively engineered by manipulating the external degrees of freedom inaugurates a novel device with potential applications in metrology, sensing and quantum simulations.\\
(Received 8 May 2021; accepted 1 June 2021; published online 8 June 2021)\\
DOI: 10.1088/0256-307X/38/7/073201
\end{abstract}

\maketitle
\end{CJK*}

 The coherent manipulation of quantum systems using periodic modulations, also known as Floquet engineering (FE), is becoming a central paradigm for the realization of synthetic quantum states and Hamiltonians \cite{Rudner_2020,Eckardt_2017,Bukov_2015}. FE has been demonstrated in a variety of different platforms including ultra-cold atoms \cite{Rudner_2020}, photonics \cite{Rechtsman_2013} and superconducting qubits \cite{Roushan_2017}. In the context of atomic gases trapped in driven optical lattices \cite{Rudner_2020}, the periodic modulation has been employed to renormalize the Hubbard Hamiltonian. In particular, the dynamical modulation can be recast into an effective tunable tunneling  \cite{Lignier_2007}. This has led to dynamical control of superfluid-insulator phase transitions \cite{Zenesini_2009}, artificial gauge fields \cite{Struck_2011,Gorg_2018,Struck_2013} and topological lattice models \cite{Cooper_2019,Aidelsburger_2013,Miyake_2013,Jotzu_2014}. 

In this work, we Floquet engineer an optical lattice clock (OLC) which is among the most accurate precision measurement devices \cite{Kolkowitz_2016,Norcia_2017,Katori_2003,Cirac_2012}. It sets the ground for next standard of time and several proposals aim to exploit their extraordinary stability and accuracy to address fundamental problems ranging from the measurements of physical constants \cite{Bloch_2012} to the detection of gravitational waves \cite{Gross_2017,Smerzi_2018}. OLC consists of an optical local oscillator stabilized by an appropriately chosen two energy levels transition of atoms trapped in a stationary lattice potential \cite{McGrew_2018}. Usually, a lattice field can significantly modify the transition energies. In OLC, this problem is addressed by engineering the magic-wavelength transition of $^{87}$Sr atoms that is well known to be insensitive to the external trap thanks to a first-order light shift cancellation \cite{Katori_2003}. Considering the high accuracy and long life time of excited clock state, OLC becomes an ideal quantum simulator in several areas, such as spin orbit coupling and SU(N) Hubbard model \cite{Zhai_2020,Ye_2017}. However, FE has not been explored so far in OLC platforms. Here we experimentally demonstrate Floquet clock bands created by a dynamical periodic modulation of the incident lattice laser frequency which is different from taking laser power as driving parameter for measuring trap frequency.

\begin{figure}[t!] 
	\centering
	\includegraphics[width=1\linewidth]{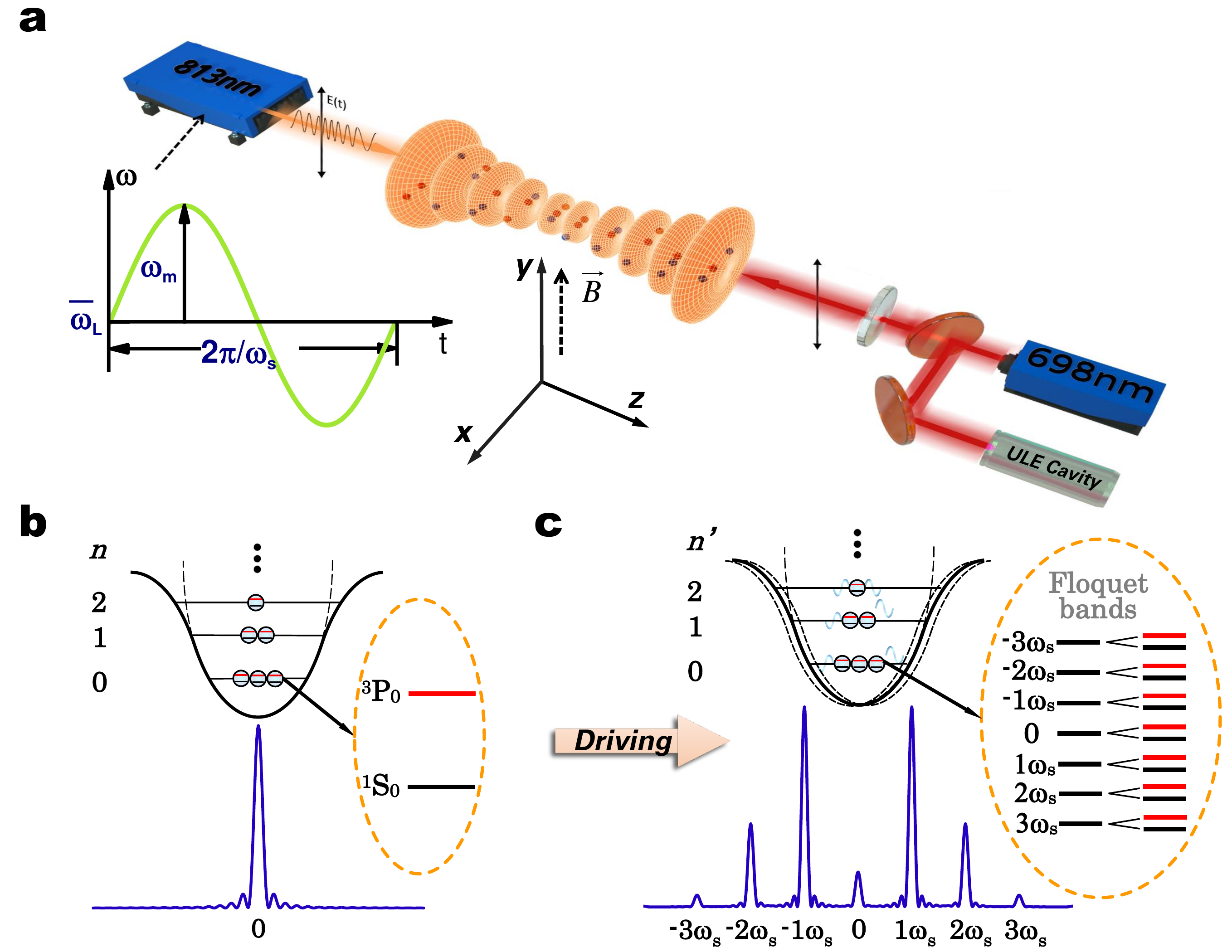}
	\caption{(a) The optical lattice potential is made by a counter-propagating laser ($\lambda_{\rm L}=813$ nm) along the $z$ direction. The beam waist ($W_0=50$ $\mu$m) locates the center of the magneto-optical trap (MOT) and its distance from the high-reflecting mirror is $L\simeq 0.3$ m. The clock laser ($\lambda_{\rm p}=698$ nm) is locked to the ultralow-expansion (ULE) glass cavity and nearly aligned along the same direction of the lattice laser to excite the clock transition.  The Floquet modulation (FM) of the lattice laser is controlled with a build-in piezo actuator. (b) At magic-wavelength $\lambda_L$ only one sharp peak at the clock transition is observed in the Rabi spectroscopy (blue curve). (c) When turning on the periodic driving, the carrier peak is split into several Floquet bands as observed in the Rabi spectroscopy (blue curve) and schematically shown in the oval inset.} \label{fig1}
\end{figure}

By trapping the atoms at the magic wave length, we show that the atomic pseudo-spin half [corresponding to the two Zeeman sublevels states $(5s^{2})^{1}$S$_{0} (|g\rangle)$ and $(5s5p)^{3}$P$_{0}(|e\rangle)$ of $F=9/2$] population dynamics is governed by a periodically driven Landau-Zener-Stuckelberg-Majorana Hamiltonian (LZSM)\cite{Sillanpaa_2006,Shevchenko_2010}. We resolve several Floquet quasi-energy bands with ultra-sensitive Rabi spectroscopy. The intensities of each band are independent on driving frequency. The number and the shape of the resonance peaks are controlled by an opportunely tailored multi-frequency driving. We study the spectroscopic sensitivity of each band by measuring the Fisher information, which provides the lower bound in sensitivity in parameter estimation theory \cite{Smerzi_ebook,Giovannetti_2011}. We show the sensitivity is not strongly reduced by the Floquet shaking potential demonstrating that, on the time scales of our experiments, dynamically induced decoherences and thermalization effects can be ignored. Our work opens to the design of a new generation of devices for sensing, metrology and quantum simulations where the internal energy structure of an optical lattice clock can be engineered by manipulating the motional atomic degrees of freedom.

\textit{Experiment Setup.}  Approximately 10$^{4}$ fermionic $^{87}$Sr atoms are cooled down to 3 \ $\mu$K and loaded in a quasi
one-dimensional optical lattice aligned with the $z$ axis. The lattice is created by a counter-propagating laser beam at magic-wavelength $\lambda_{\rm L} =$ 813 nm \cite{Takamoto_2003,Takamoto_2005} (Fig. \ref{fig1}a), so that the atoms at the dipole-forbidden transition energy levels  $|g\rangle$ and $|e\rangle$ feel the same lattice potential \cite{Katori_2003} (Fig. \ref{fig1}b). Meanwhile, the atom is prepared at the Zeeman sublevel of $m_F=+9/2$. We load about one thousand lattice sites separated by barriers of height $V_{0}/E_{r} \approx$ 90 ($E_{r}$ is the recoil energy) which hinder inter-site tunneling. We Floquet engineer our system by periodically driving the piezo actuator \cite{SM}.

The frequency of the lattice laser is modulated as $\omega_{\rm L}(t) = {\bar{\omega}}_{\rm L} + f(t)$, where ${\bar{\omega}}_{\rm L} = 2\pi c/\lambda_{\rm L}$ is the average lattice carrier frequency and
\begin{equation}
f(t)=\sum_{m=1}^{N} \omega_{m}\sin(\omega_{\rm s} mt)
\label{eq1}
\end{equation}
is a $T= 2 \pi/\omega_{\rm s}$ periodic $N$-modes function. We first consider a monochromatic driving $f(t)=\omega_{1}\sin(\omega_{\rm s}t)$, while multi-mode will be discussed later. The intensity of lattice laser along $z$ direction becomes $I=I_{0}\sin^{2}\left(\bar{\omega}_{\rm L}(z+\int v(t) dt)/c\right)$, where $v(t)\simeq \omega_{1}\omega_{\rm s}L \cos( \omega_{\rm s}t)/\bar{\omega}_{\rm L}$ is an effective lattice velocity. In the lattice co-moving frame, the frequency of optical clock laser (CL) $\omega_{\rm p}$ is shift to $\omega_{\rm p}'(t)=(1-v(t)/c)\omega_{\rm p}$ due to the relativistic Doppler effect \cite{SM}.

\textit{Model. } When internal clock states being mapped to spin-$1/2$ and described by Pauli matrix $\hat{\sigma}$, the Hamiltonian in the co-moving frame of a single atom interrogated by the clock laser and trapped in a driven periodic potential can be written as
$\widehat{H} = {\widehat{H}}_{\text{ext}} + {\widehat{H}}_{\text{LZSM}} + {\widehat{H}}_{\text{c}}$ \cite{SM},

where
\begin{eqnarray}
\nonumber
{\widehat{H}}_{\text{ext}} & = & [ \frac{{\widehat{p}}^{\text{2}}}{\text{2}M}  + \frac{\alpha_{0}I_{0}}{\text{2}\epsilon_{\text{0}}c}e^{- 2r^{2}/W_{0}^{2}}\text{sin}^{\text{2}}\left( \frac{{\bar{\omega}}_{\rm L}}{c}z \right)\\
&& - \frac{M\omega_{1}\omega_{\rm s}^{\text{2}}L}{{\bar{\omega}}_{\rm L}}z\text{sin}{(\omega}_{\rm s}t)]{\widehat{\sigma}}^{(0)},
\label{eq2}
\end{eqnarray}
governs the motion of atom and
\begin{equation}
{\widehat{H}}_{\text{c}} = \frac{\alpha_{c}I_{\text{0}}}{4\epsilon_{\text{0}}c}\sin\left( \omega_{\rm s}t \right){\text{sin}}^{\text{2}}\left( \frac{{\bar{\omega}}_{\rm L}}{c}z \right){\widehat{\sigma}}^{\left( \text{3} \right)}
\label{eq3}
\end{equation} 
is coupling Hamiltonian provided by a spin-dependent optical lattice potential (notice that this term has a purely dynamical origin, while spin-dependent periodic potentials have been created with polarized standing waves laser fields before\cite{Mandel_2003,Dai_2016}). The parameter $\alpha_{c}$ is proportional to the difference between polarizability derivatives of the two spin states calculated at the magic-wavelength \cite{SM}.

Therefore, the value of coupling constant $\frac{\alpha_{c}I_{0}}{4\epsilon_{0}c} \approx 1$\,Hz is so weak that Eq.\,(3) can be neglected on time scales of the order of a second. This is of the same order of the dephasing time of our optical lattice clock caused by the finite temperature of the atomic sample. The spin dynamics of the atomic gas is governed by the Landau-Zener-Stuckelberg-Majorana Hamiltonian \cite{Shevchenko_2010}:
\begin{equation}
{\widehat{H}}_{\text{LZSM}}\left( t \right)  =  \frac{\hbar}{\text{2}}\left( \delta + \omega_{\rm p}\frac{v\left( t \right)}{c} \right){\widehat{\sigma}}_{\boldsymbol{n}}^{\left( \text{3} \right)} + \frac{g_{\boldsymbol{n}}}{\text{2}}{\widehat{\sigma}}_{\boldsymbol{n}}^{\left( \text{1} \right)}
\label{eq4} 
\end{equation}
where \(\delta = \omega_{\text{0}} - \omega_{\rm p}\) is the detuning, \(\omega_{\text{0}}\) is the clock transition frequency of \textsuperscript{87}Sr, \(g_{\boldsymbol{n}}\) is an effective coupling strength of the atoms with CL. Notice that the spatial
driving enters as an effective modulation proportional to \(v\left( t \right)\) while the external degrees of freedoms of the
atoms remain unchanged when ignore the small linear potential term in Eq. \ref{eq2}: $\hat{\sigma}_{\boldsymbol{n}}$ describes the internal degrees of freedom for atoms at external eigenstates $\boldsymbol{n} = (n_{z},n_{r})$ with the eigenenergies \(E_{\boldsymbol{n}}/h = \nu_{z}(n_{z} + \text{1/2}) + \nu_{r}(n_{r} + \text{1})\) of the trapping potential having longitudinal and transverse trap frequencies \(\nu_{z} =\) 64.8 kHz and \(\nu_{r} =\) 250 Hz, respectively.

\begin{figure}
	\centering
	\includegraphics[width=0.5\linewidth ]{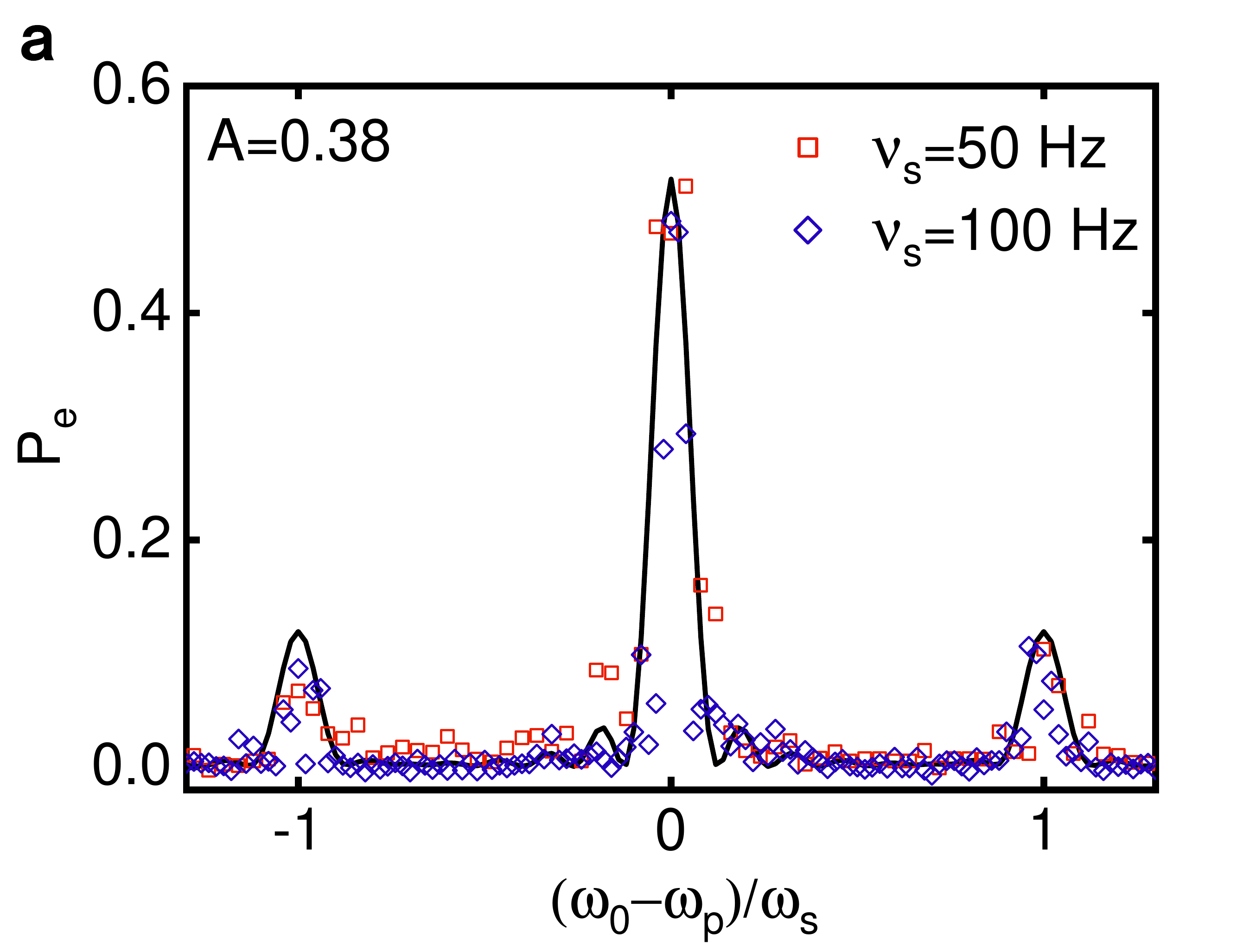}\hfill
	\includegraphics[width=0.5\linewidth]{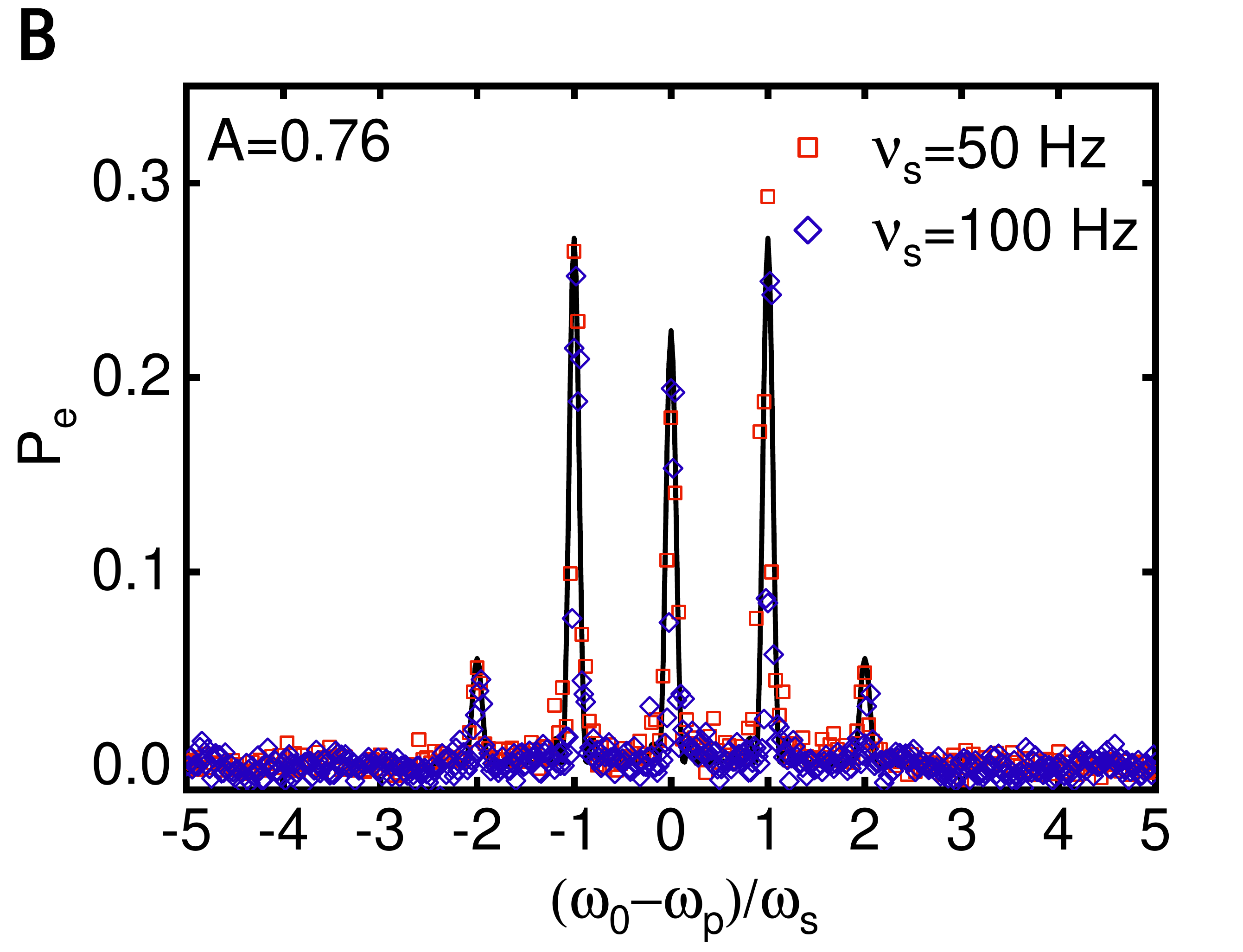} \\
	\includegraphics[width=0.5\linewidth ]{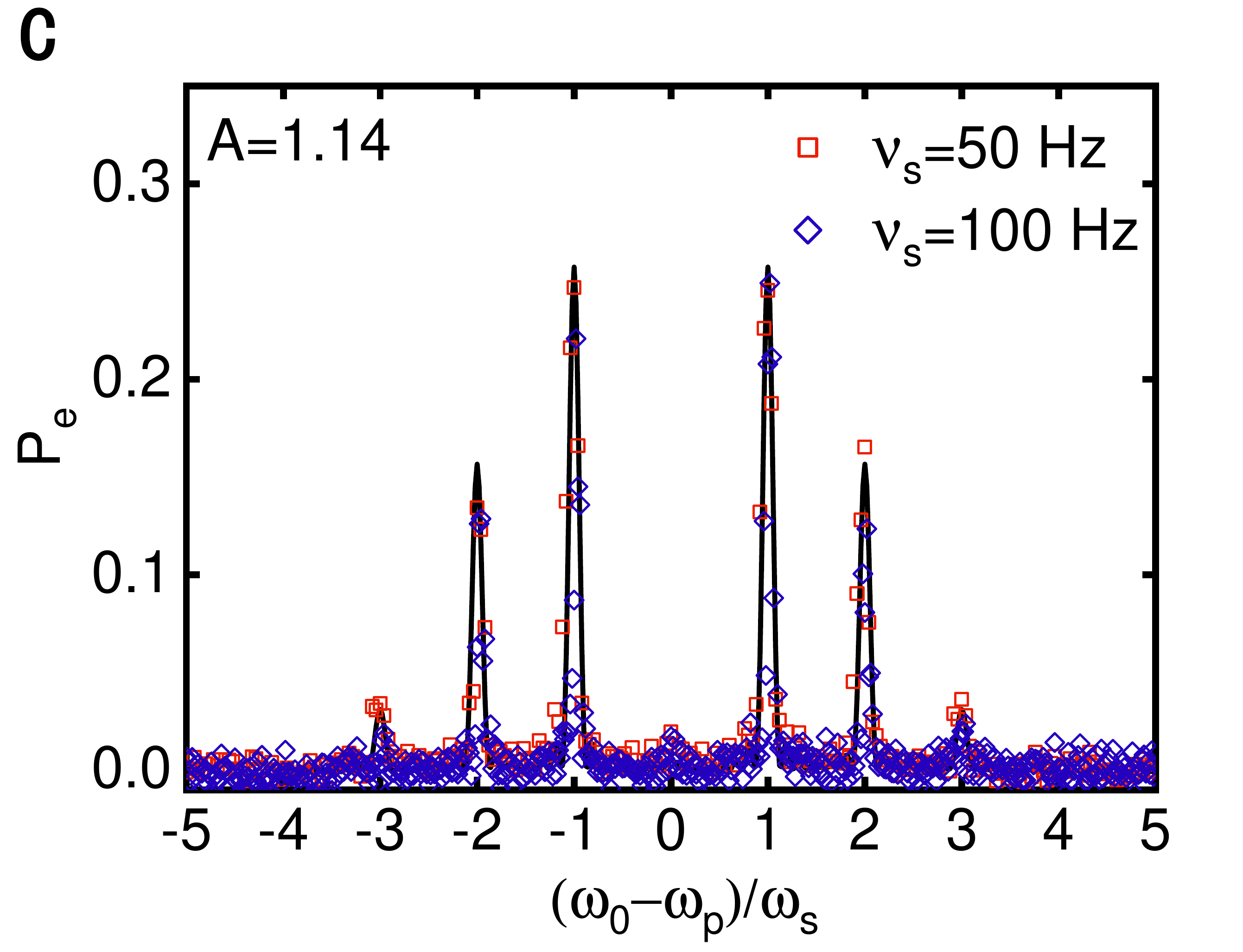}\hfill
	\includegraphics[width=0.5\linewidth]{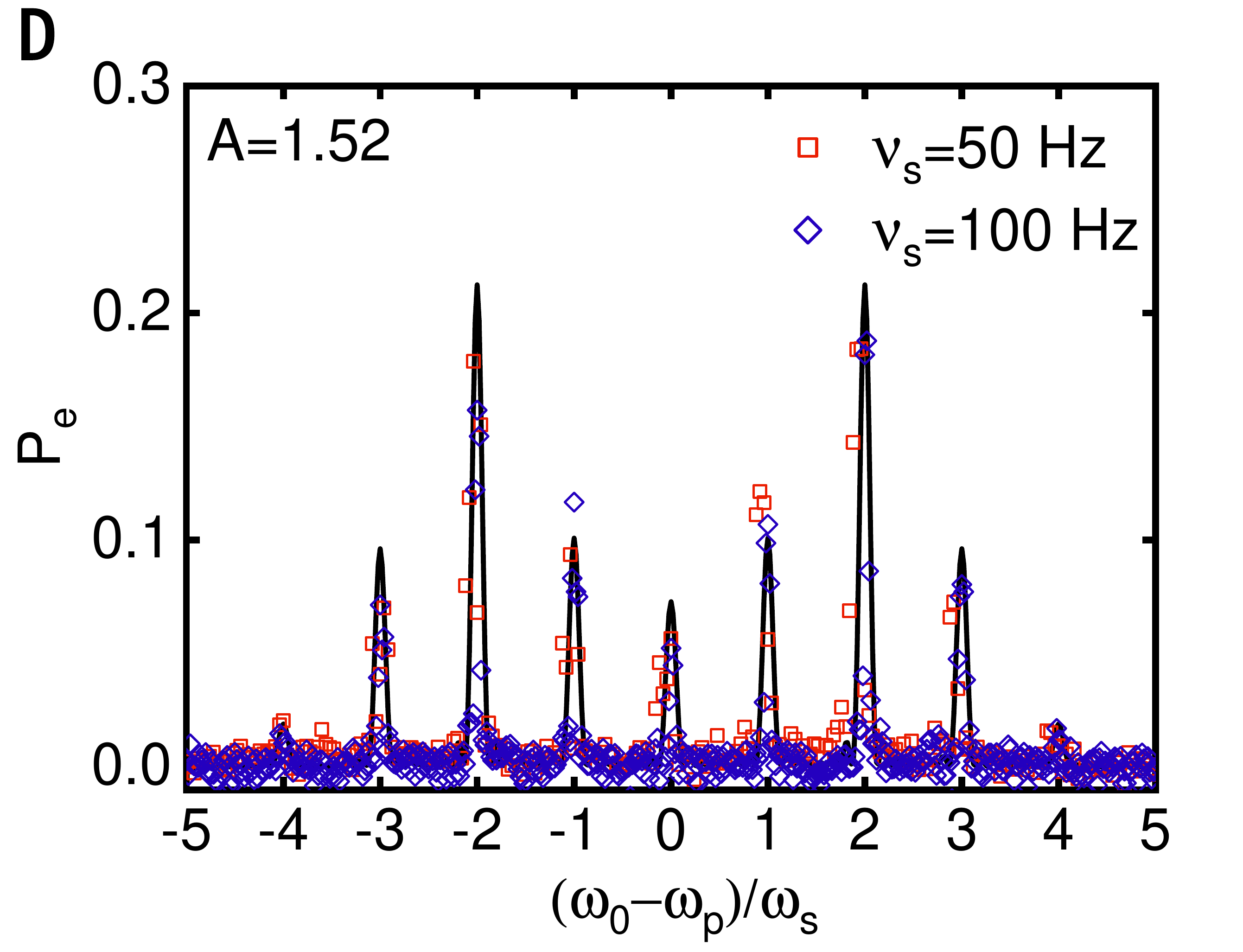}\\
	\includegraphics[width=0.5\linewidth ]{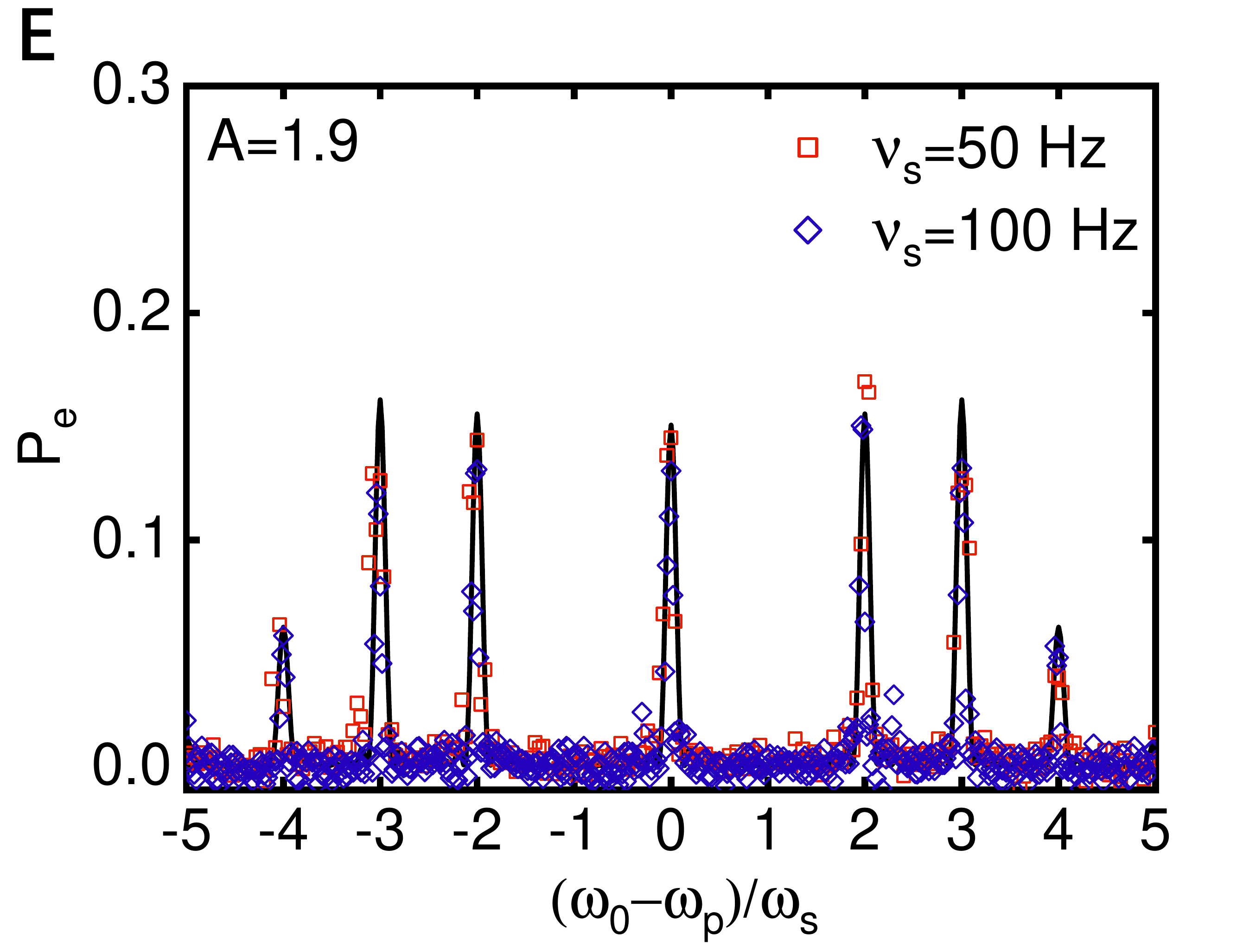}\hfill
	\includegraphics[width=0.5\linewidth ]{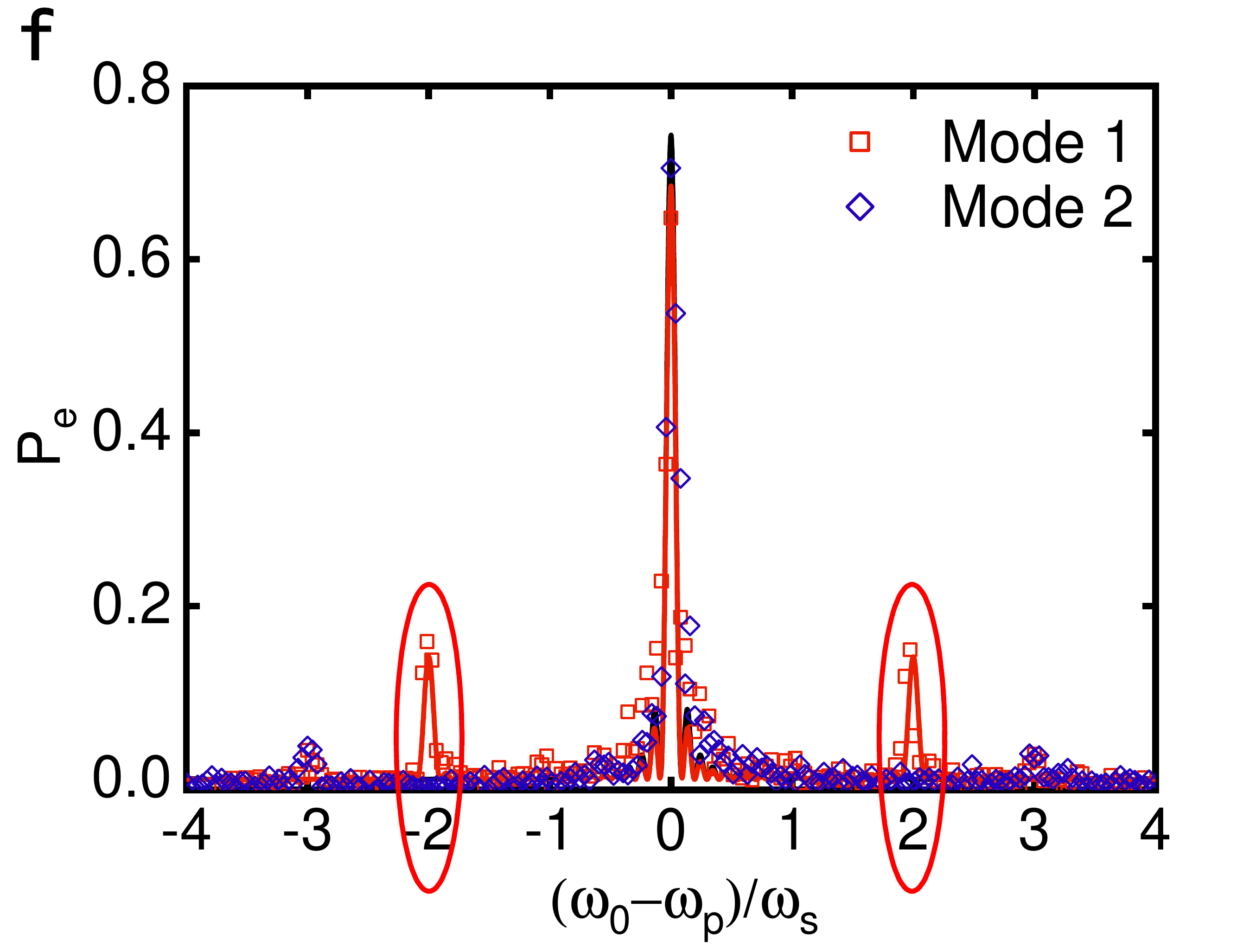}
	\caption{Rabi spectroscopy of the Floquet bands. (a)-(e) Monochromatic driving with different amplitudes $A$. The experimental measurements at the probe time $t_{\rm p}=150$ ms and driving frequencies $\nu_{\rm s}=\omega_{\rm s}/2\pi=50$ Hz (red square) and $\nu_{\rm s}=\omega_{\rm s}/2\pi=100$ Hz (blue diamond) are compared with theoretical results Eq.(\ref{eq5}) (solid black line). (f) Three-frequencies driving with two sets of coefficients: $\boldsymbol{A}_1=\{0.065,0.345,0.243\}$(mode 1) and $\boldsymbol{A}_2 = \{0.005,0.005,0.16\}$ (mode 2). The measured Rabi spectra of mode 1 (red square) and mode 2 (blue diamond) are presented. In both cases, the probe time is $t_{\rm p}=200$ ms and the driving frequency $\omega_{\rm s}/2\pi=50$ Hz.  In red ovals, the suppression of second order FBs when using the Mode-2 set of parameters is shown, demonstrating the possibility of manipulating specific FB. The solid red and black lines are the theoretical Rabi spectrum of mode 1 and mode 2, respectively.} \label{fig2}
\end{figure}

\begin{figure}[t!]
	\centering
	\includegraphics[width=0.5\linewidth ]{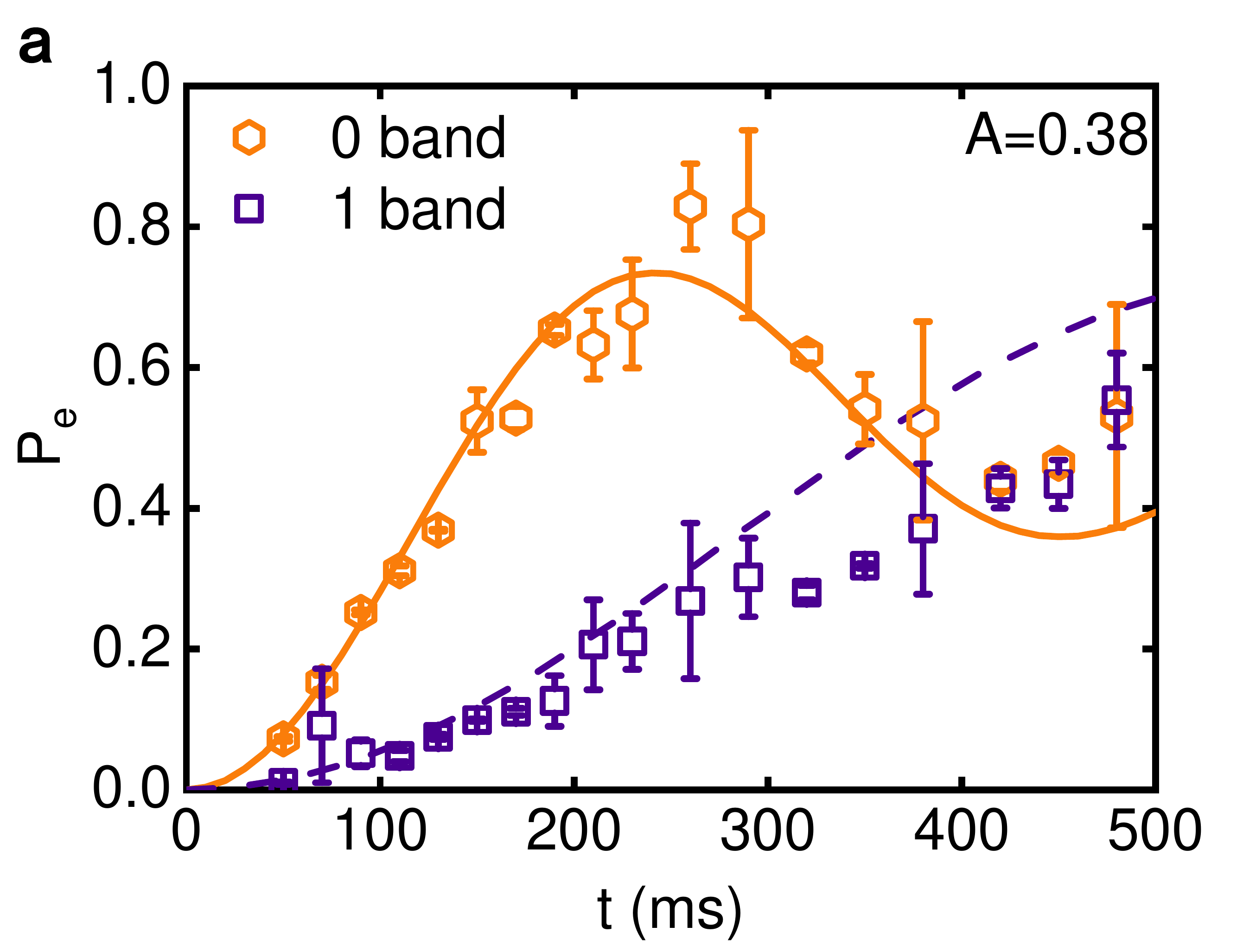}\hfill
	\includegraphics[width=0.5\linewidth]{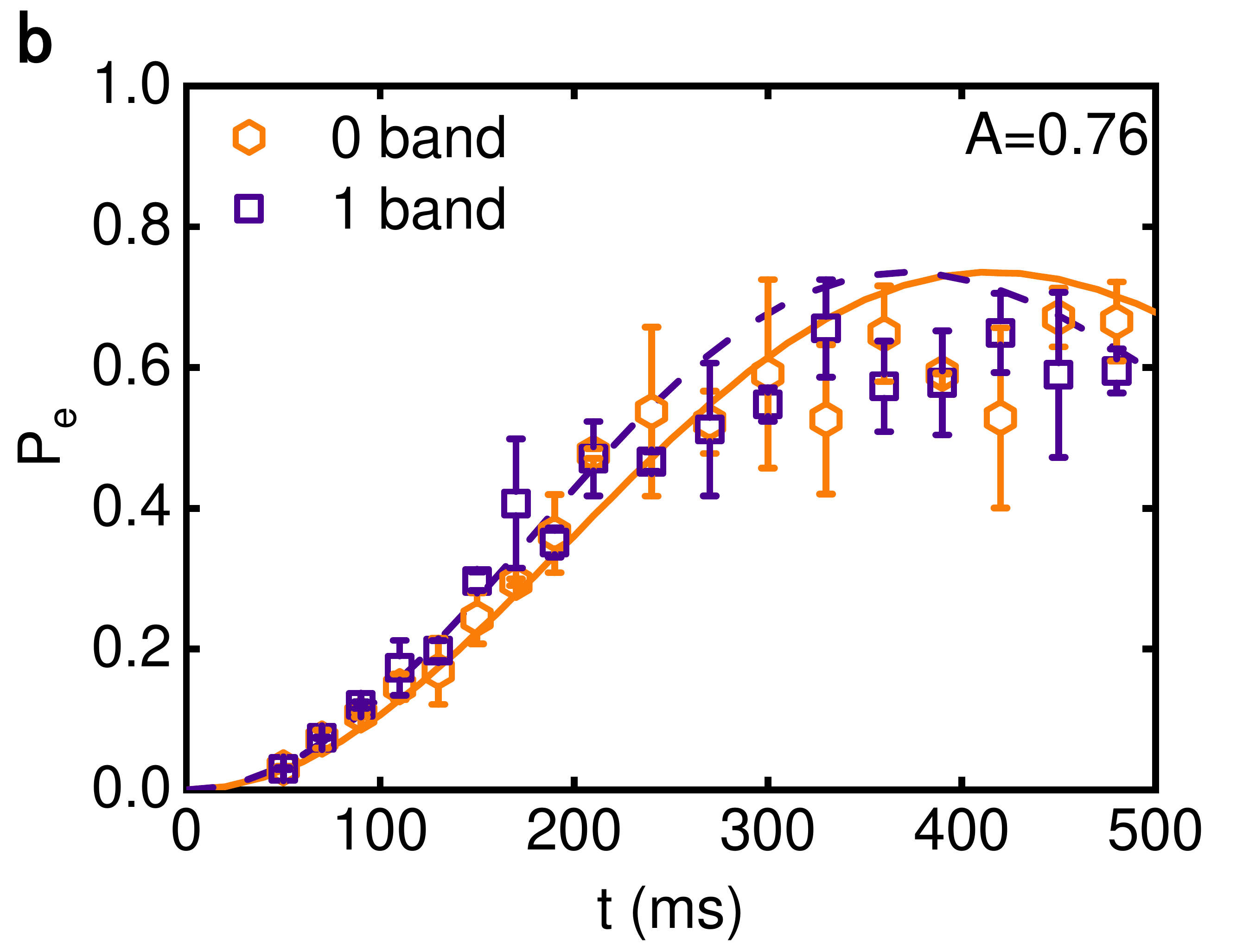}\\
	\includegraphics[width=0.5\linewidth ]{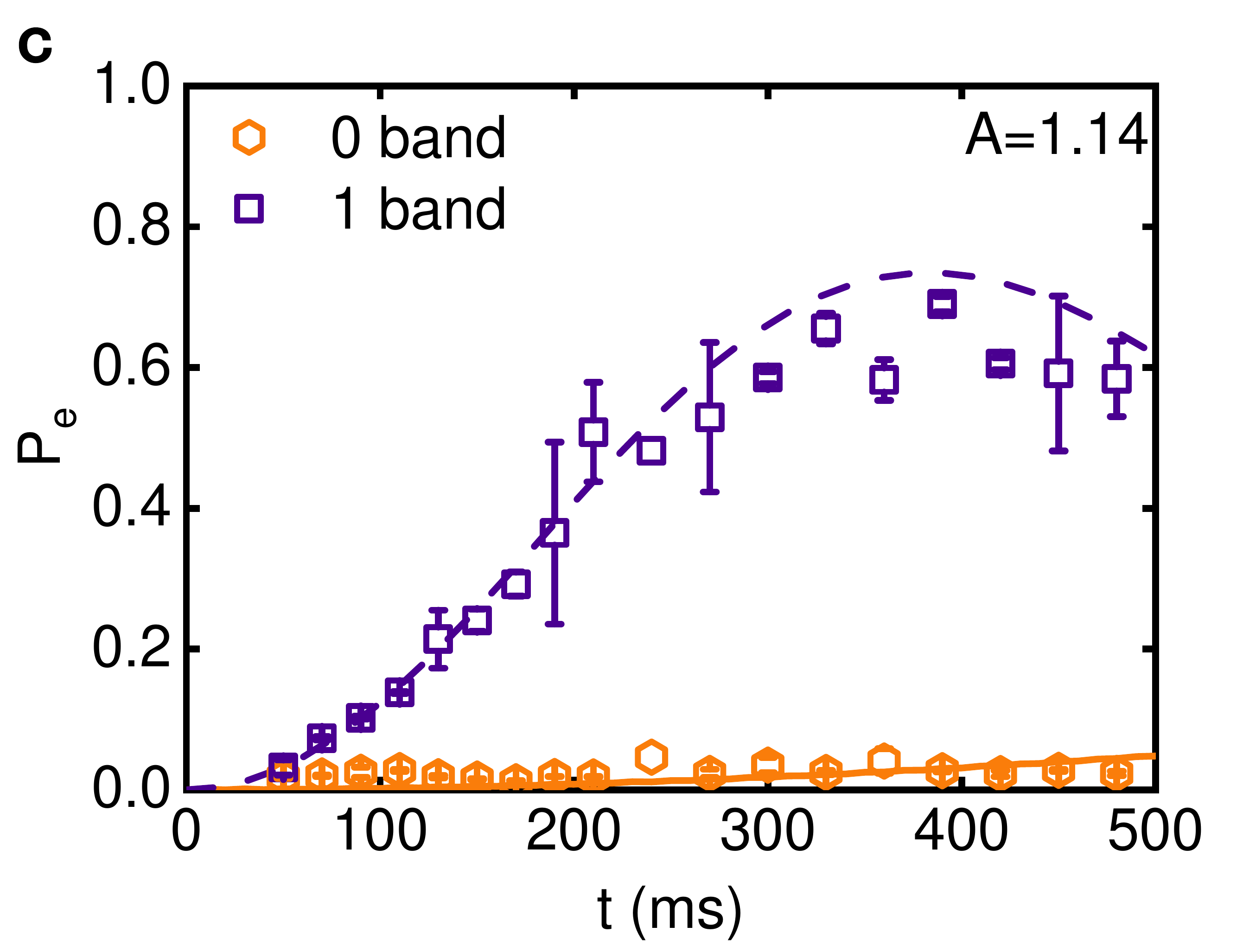}\hfill
	\includegraphics[width=0.5\linewidth ]{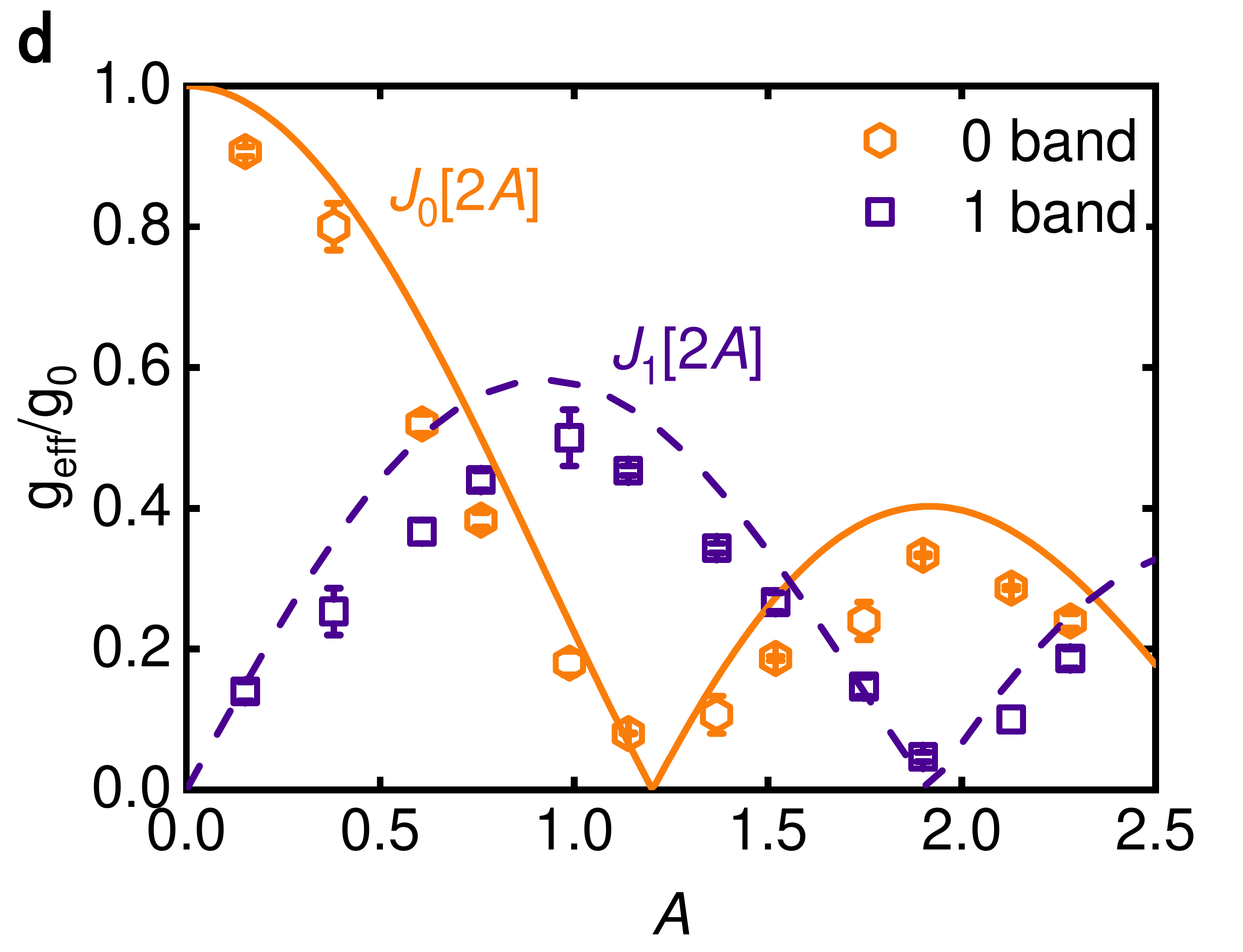}\hfill
	\caption{Theoretical (solid lines) and experimental (dashed lines)
		Rabi oscillations at driving amplitudes (a) $A=0.38$, (b) $A=0.76$ and (c) $A=1.14$. The orange hexagons
		are for the zeroth band while the blue squares for the first band. (d) The ratio between Floquet modulated Rabi frequencies with Rabi frequencies of the zeroth and first bands measured experimentally are compared with the Bessel function  predicted by the Floquet theory.} \label{fig3}
\end{figure}

\textit{Rabi spectroscopy. } The measurements of atomic energies are performed by high-precision clock Rabi spectroscopy that operates at a fractional instability of 10\textsuperscript{-15} \cite{Wang_2018}. The spectroscopic CL is locked to an ultralow-expansion cavity having a linewidth of approximately 1 Hz. A slight unavoidable misaligning between CL and lattice axis, see Fig. \ref{fig1}a, induces a coupling to the suppressed radial motional modes and a correction arising from the thermal distributions of the coupling strength:
$$g_{\boldsymbol{n}} = g_{\text{0}}e^{- (\eta_{z}^{\text{2}} + \eta_{r}^{\text{2}})/\text{2}}L_{n_{r}}(\eta_{r}^{\text{2}})L_{n_{z}}(\eta_{z}^{\text{2}}),$$
where \(g_{0}/h = \)3.3 Hz, and \(L_{n}\) is the \(n\)th order Laguerre polynomial with Lamb-Dick parameters \(\eta_{z} = \sqrt{h/(\text{2}M\nu_{z})}/\lambda_{\rm p}\) and \(\eta_{r} = \sqrt{h/(\text{2}M\nu_{r})}\delta\theta/\lambda_{\rm p}\) with $\delta\theta$ being the misaligned angle between the lattice and the clock laser. At the end of each spectroscopic probe, the number of atoms in the \(|g\rangle\) and \(|e\rangle\) states are determined by using a cycling transition \cite{SM}. This provides the normalized population fraction \(P_{e}\) of the \(|e\rangle\) state. After many repetitions of the measurements, data are collected while the CL is scanned across the clock transition to eventually construct the Rabi spectrum (Fig. \ref{fig2}). Rabi oscillations as a function of the probe pulse time \(t_{\rm p}\) are shown in Fig. \ref{fig3}. Both the Rabi spectroscopy and the Rabi oscillations are performed while modulating the system with Floquet periodic driving. Notice that since the measurement processes last only hundreds of milliseconds, we can ignore the spontaneous emission due to the long lifetime of the excited states. According to the Floquet theory \cite{Eckardt_2017}, the clock energy levels are split to several Floquet bands (FB) as depicted in Fig. \ref{fig1}c. The experimental results of the Rabi spectroscopy obtained at different driving modulation amplitudes \(\omega_{1}\) and frequencies \(\omega_{\rm s}\) are presented in Fig. \ref{fig2}a-e.

We observe, in particular, that (i) sharp FBs are separated by intervals \(\omega_{\rm s}\), each band having line width of a few Hz ; (ii) the number of FBs is increasing with the renormalized driving amplitude \(A = \omega_{\text{1}}\omega_{\rm p}L/\text{2}{\bar{\omega}}_{\rm L}c\); (iii) the intensity of Floquet Rabi spectra depends on the values of \(A\) [not on \(\omega_{\rm s}\)] after rescaling the detuning as \((\omega_{\text{0}} - \omega_{\rm p})/\omega_{\rm s}\). These phenomena can be quantitatively understood within the Floquet theory \cite{Eckardt_2017}. Because of the ultra-stable narrow optical CL and \(\delta \ll \text{2}\pi\nu_{z}\), the Rabi oscillations mainly occur within a fixed external quantum numbers \(\boldsymbol{n}\). We can therefore study the dynamical evolution of the spin populations in an extended Hilbert space, consisting of the direct product of the original spin and Floquet quasi-levels \(\lbrack|g\rangle,|e\rangle\rbrack\bigotimes\left\lbrack 1,e^{\pm i\omega_{\rm s}t},e^{\pm \text{2}i\omega_{\rm s}t},e^{\pm \text{3}i\omega_{\rm s}t}\text{...} \right\rbrack\). In the resonance region \(\omega_{\rm s} \gg g_{n_{z},n_{r}}\) \cite{Shevchenko_2010}, we define an effective Rabi frequency for the \(k\)th FB as \(g_{\boldsymbol{n}}^{k} = g_{\boldsymbol{n}}J_{k}\lbrack\text{2}A\rbrack\), where \(J_{k}\lbrack\text{2}A\rbrack\) is the \(k\)th order first kind Bessel function. Thus, the excited state population for \(k\)th FB is
\begin{equation}
P_{e}^{k}(\delta,t)=\sum_{\boldsymbol{n}}q_{z}(n_z)q_{r}(n_r)\left(\frac{g_{\boldsymbol{n}}^{k}}{\hbar R_{k}}\right)^2 \sin^{2}\left[\frac{R_{k}}{2} t\right],
\label{eq5}
\end{equation}  
where \(R_{k} = \sqrt{(g_{\boldsymbol{n}}^{k}/\hbar)^{\text{2}} + (\delta - k\omega_{\rm s})^{\text{2}}}\). The \(q_{z}(n_{z})\) and \(q_{r}(n_{r})\) are the statistical distributions of the atoms among eigen-energies of the lattice trapping potential \cite{Blatt_2009}. It is important to notice here that the atomic statistical distributions in periodically driven systems are, in general, not fully understood in the literature. We could still expect a Boltzmann distribution with an effective temperature since, in our case, the Floquet gap \(\hbar\omega_{\rm s}/k_{B}\) [\(k_{B}\) is the Boltzmann constant] is about a few nK and is much smaller than the longitudinal trap gap energy \(h\nu_{z}/k_{B}\) which is several \(\mu\)K. We therefore assume the Boltzmann distributions  \(q_{z(r)}[n_{z(r)}] = [1 - Z_{z(r)}]Z_{z(r)}^{n_{z(r)}}\) with \(Z_{z(r)} = e^{- h\nu_{z(r)}/[k_{B}T_{z(r)}]}\), where the longitudinal and transverse trap effective temperatures \(T_{z} =\) 2.96 \(\mu\)K and \(T_{r} =\) 3.68 \(\mu\)K are extracted by the experimental side bands spectra \cite{Blatt_2009}.

The good agreement between the experimental spectra reported in Fig. \ref{fig2}a-e and the theoretical predictions of Eq. \ref{eq5} demonstrates the validity of Boltzmann statistics assumption. Furthermore, from Eq. \ref{eq5} we gather that the center of Rabi spectrum \(P_{e}^{k}(\delta,t_{\rm p})\) is provided by \(\omega_{\rm s}\) while the line shapes mainly depend on \(A\), in agreement with what observed in (i) and (iii). The point (ii) is the direct consequence of the Bessel functions modulations of the Rabi frequency. By increasing the renormalized driving amplitude \(A\), higher order Bessel functions become relevant and an increasing number of FBs emerges. At the same time, the weights of a few of them may decrease or be totally suppressed, as can be observed in the zeroth band at \(A =\) 1.14 and in the first FB at \(A =\) 1.9 in Fig. \ref{fig2}. In Fig. \ref{fig3}a-c we show the Rabi oscillations at different values of the strength \(A\). Notice that at \(A =\) 1.14, the zeroth band is totally eliminated, as also evident from Fig. \ref{fig2}c. We can probe the height \(P_{e}^{k}(\delta,t_{\rm p})\) of \(k\)th FB at a time \(t_{\rm p}\) and compare it with the non-driven case at a different time $t_{\rm p}'$, defined as the time when the values of the two peaks are the same. The ratio between Floquet modulated and natural Rabi frequencies is \(g_{\text{eff}}^{k}/g_{\text{0}} = t_{\rm p}'/t_{\rm p} = J_{k}\lbrack\text{2}A\rbrack\). As shown in Fig. \ref{fig3}d, this Bessel function dependence emerges quite clearly in the experimental results.

We now extend the monochromatic driving to $N$ multi modes periodical function, see Eq.\ref{eq1}. In this case the $k$th Floquet level is modulated by a rather complex combination of Bessel functions :
\begin{equation}
\mathcal{J}_{k}[\boldsymbol{A}]=\sum_{\{k_1,k_2,\ldots,k_N\}}\prod_mJ_{k_m}[2A_m]
\label{eq6}
\end{equation}
where $\{\boldsymbol{k}_m\}$ is the subset of $\{k_1,k_2,\ldots,k_N\}$ with the constraint \(\sum_{m}^{}mk_{m}=k\) and \(A_{m} \equiv \omega_{m}\omega_{\rm p}L/\text{2}{\bar{\omega}}_{\rm L}c\) is the renormalized driving amplitude for $m$th mode. By appropriately choosing the values of \(A_{m}\) of each mode, we independently modulate the FBs while keeping the zeroth band nearly unchanged. As shown in Fig. \ref{fig2}f, the experimental results of three-frequencies driving are in good agreement with the theoretical prediction of Eq. \ref{eq5} after replacing \(J_{k}\lbrack A\rbrack,\) with \(\mathcal{J}_{k}\lbrack\overrightarrow{A}\rbrack\). It would also be
possible to create asymmetric distributions by introducing a different phase in each mode.

\textit{Sensitivity of the Rabi spectroscopy. } We now estimate the spectroscopic sensitivity of the modulated optical clock by measuring the Fisher information (FI). The FI plays a central role in parameter estimation theory since it determines the Cramer-Rao sensitivity lower bound \cite{Smerzi_2018,Smerzi_ebook,Giovannetti_2011}. It can also be shown that the FI is inversely proportional to the Allan variance,\cite{Rihele_ebook} \(\sigma_{\text{Al}} \sim \text{1}/(\tau F)\), where \(\tau\) is the measurement time, whenever time noise correlations of the local oscillator can be neglected. Here the parameter to be estimated is the detuning \(\delta\), with the Fisher information

\begin{equation}
F(\delta) = \frac{1}{P_e(\delta)(1-P_e(\delta))} \left( \frac{\partial P_e(\delta)}{\partial \delta} \right)^2.
\label{eq7}
\end{equation}  
\begin{figure}
	\centering
	\includegraphics[width=0.50\linewidth ]{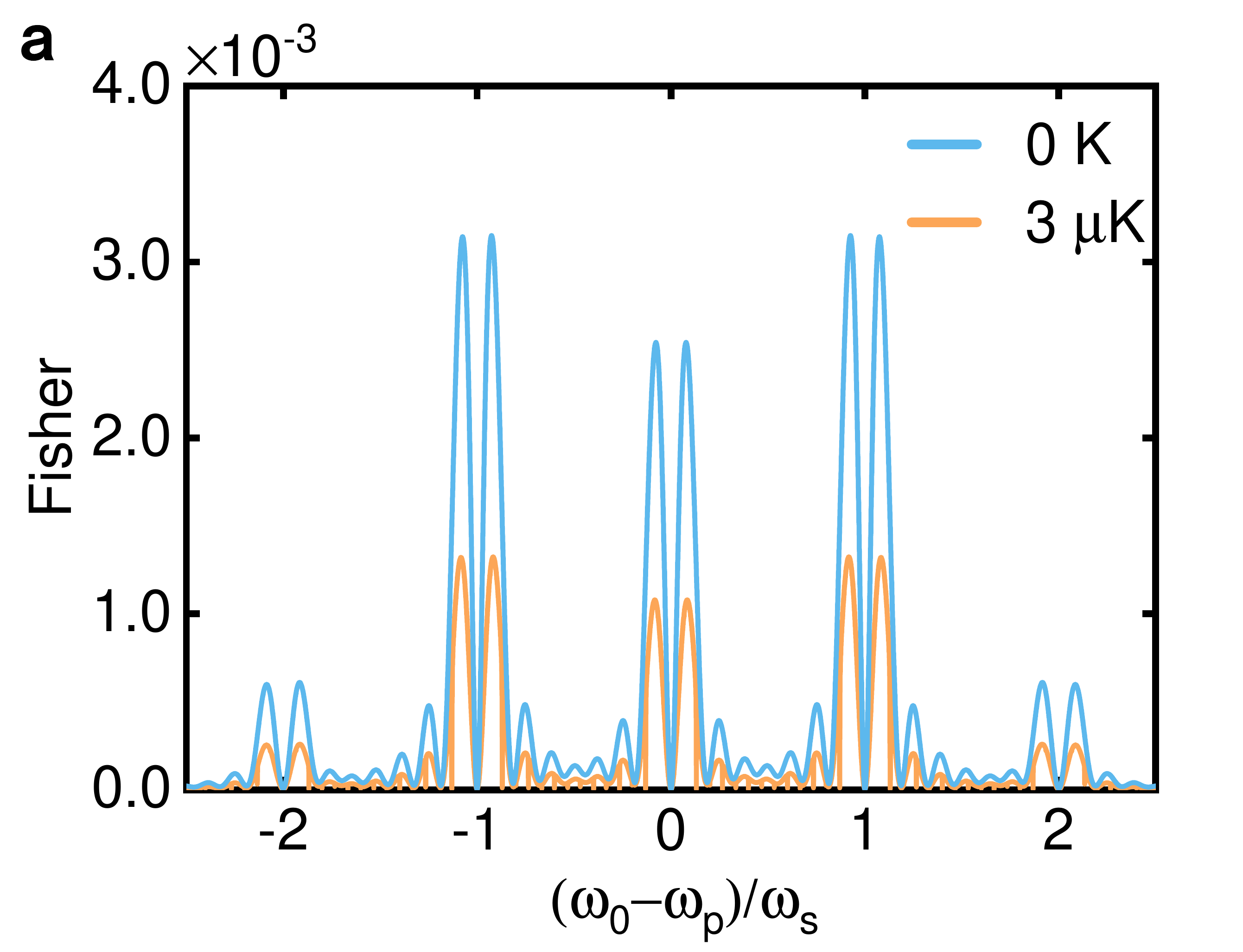}\hfill
	\includegraphics[width=0.50\linewidth]{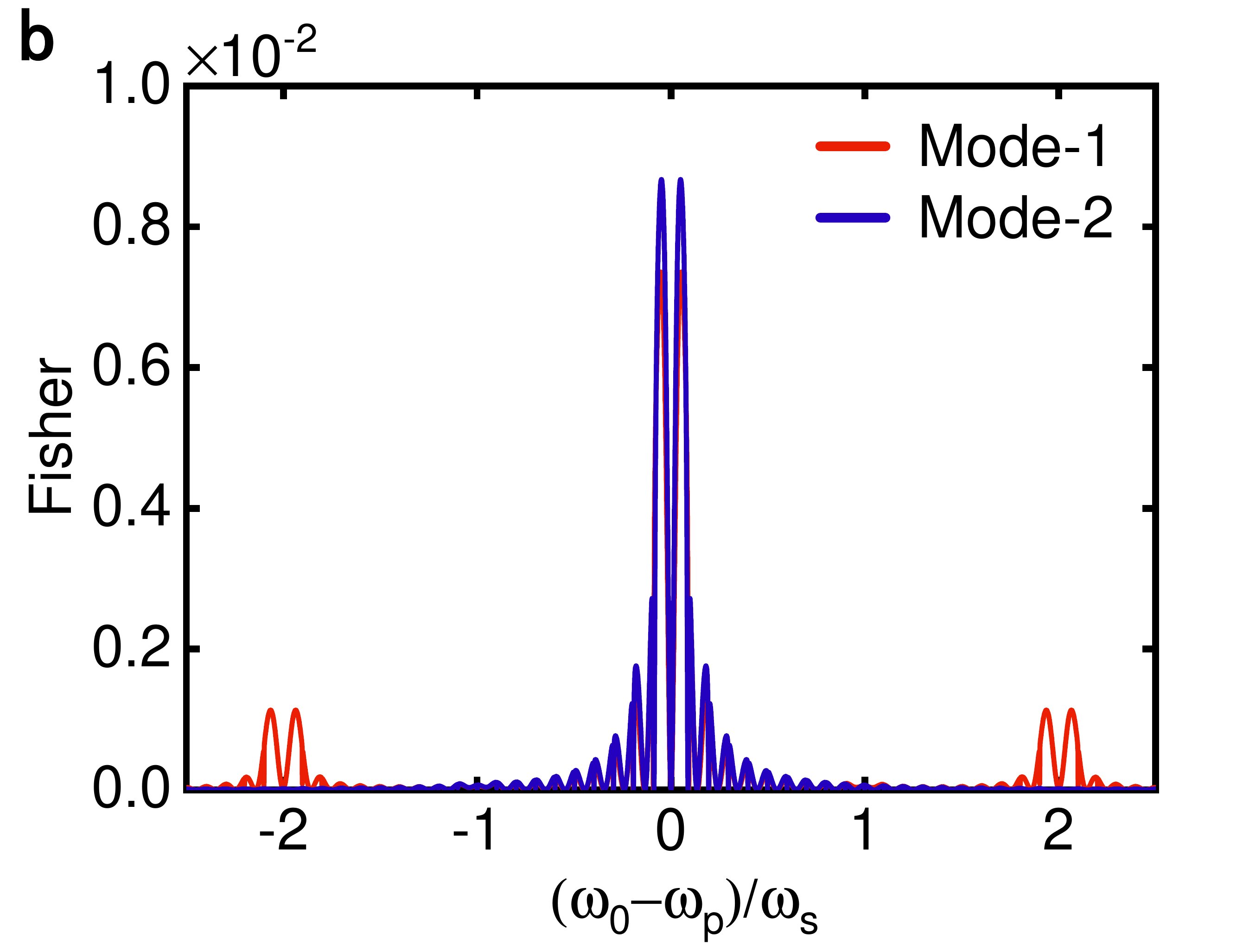}\\
	\includegraphics[width=0.50\linewidth ]{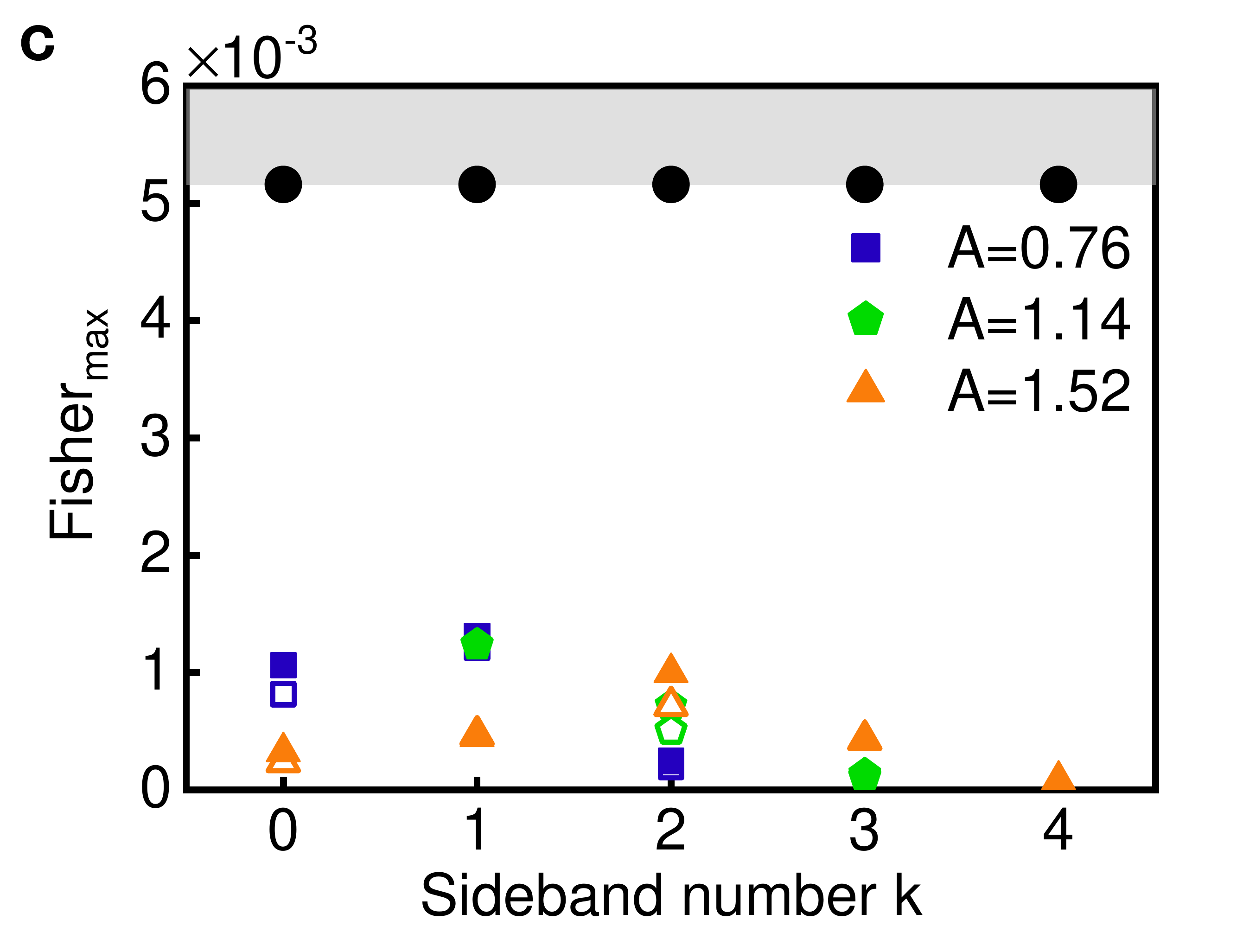}\hfill
	\includegraphics[width=0.50\linewidth ]{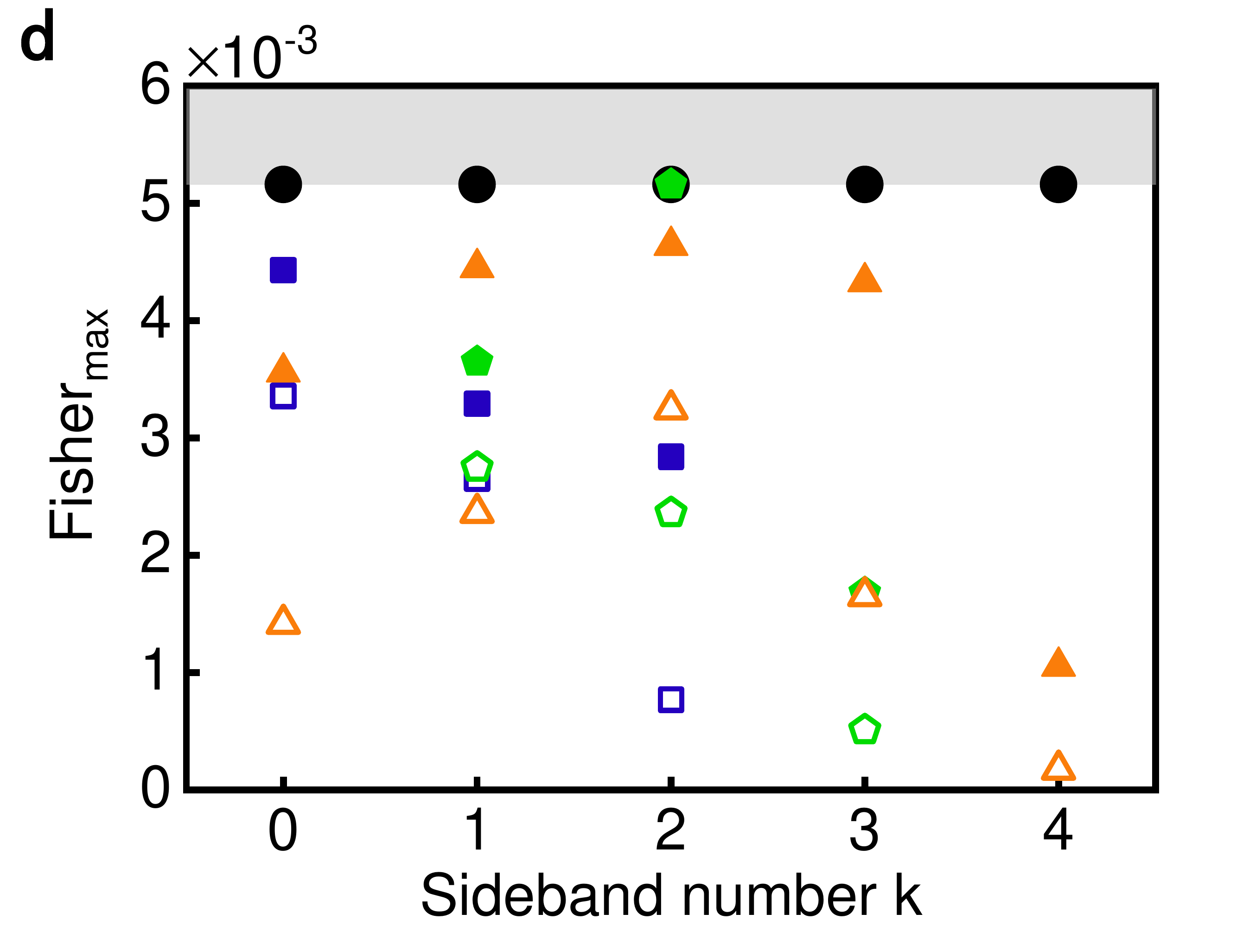}
	\caption{Fisher information of Floquet distributions. (a) The FI as a function of the rescaled detuning with parameter values as in Fig. \ref{fig2}b in the cases of zero and finite temperatures with A=0.76; (b) FI for the mode-1 and mode-2 three-frequency driving cases, see Fig. \ref{fig2}f; Experimental (hollow icons) and theoretical (solid icons) values of the maximum Fisher information in the case of (c) $g_0/h=3.3$ Hz  and (d) $g_0/h=12$ Hz. The dark filled circles are the maximum theoretical value of the Fisher information, maximized over all possible values of $g_0$, at finite temperature. (the maximum Fisher information at zero temperature is $\sim 9.1 \times 10^{-3}$ at $g_{0max}/h \simeq 3.34$ Hz while the thermal effects reduces it to $\sim 5.2 \times 10^{-3}$).}
\label{fig4}
\end{figure}
Notice that, since we have dichotomic measurements, the Fisher information coincides with the error propagation expression \cite{Itano_1993}. Given the symmetry of the probability distributions, the FI of each band has a two-peak structure, with the maximum determined by the competition between a maximum slope and a minimum fluctuation (provided by the denominator of Eq. \ref{eq7}). In Fig. \ref{fig4}a-b we show the theoretical values of the Fisher information obtained with \(P_{e}(\delta,t)\), Eq. \ref{eq5},  calculated at a fixed \(g_{0}\). In Fig. \ref{fig4}a we compare the zero and finite temperature cases. It is evident that the value of FI strongly depends on the Floquet band and, as expected, is depleted by the temperature of the atomic gas. In Fig. \ref{fig4}b we show the FI for the mode-1 and mode-2 three-frequency drive. In Fig. \ref{fig4}c-d we compare the theoretical values of the FI with the experimental results for different values of \(g_{\text{0}}\) and $A$. The experimental value of the FI is recovered with a fit of the probability distributions 
obtained from the Rabi spectra reported in Fig.  \ref{fig2}, see \cite{SM}. The agreement between the theoretical calculation and the experimental values is remarkable given that the theoretical Fisher information has been calculated with the ideal probability distributions by only taking in account temperature effects, neglecting all other source of decoherence like the laser fluctuations that are present in the experimental realizations. Furthermore, since in the un-driven case the maximum value of the theoretical FI is approximately $5.2\times10^{-3}$ the sensitivity of lower order Floquet bands is not reduced.

\textit{Conclusions and Outlook. } We have engineered an optical atomic clock by periodically driving the trapping lattice potential, resolved with Rabi spectroscopy several Floquet quasi-energy bands. Also, we have demonstrated the possibility to selectively manipulate a chosen band by appropriately adjusting the driving amplitudes of different driving modes. In future work, it will be possible to shape the inter-well tunneling barriers of the lattice by modulating the potential of Eq.  \ref{eq2} as already demonstrated in \cite{Lignier_2007,Zenesini_2009,Struck_2011,Gorg_2018,Struck_2013,Cooper_2019,Aidelsburger_2013,Miyake_2013,Jotzu_2014}. The spin-lattice coupling Hamiltonian Eq.  \ref{eq3} can be switched on to introduce entanglement between spatial and spin coordinates. Furthermore, it will be possible to control the inter-atomic interaction in bosonic clocks with Feshbach resonances \cite{Chin_2010}. These explorations will open to the possibility of creating a novel generation of quantum simulators with the experimental measure of the Fisher information witnessing multiparticle entanglement \cite{Pezze_2009,Pezze_2016}.

\textit{Acknowledgments} We thank A. Bertoldi, L. Pezz\`{e} and N. Poli are acknowledgedfor helpful discussions.This work is supported by the National Natural Science Foundation of China (Grant Nos. 61775220, 11804034, 11874094, 12047564, 11874247, 11874246), the Key Research Project of Frontier Science of the Chinese Academy of Sciences (Grant No. QYZDB-SSW-JSC004), and the Strategic Priority Research Program of the Chinese Academy of Sciences (Grant Nos. XDB21030100 and XDB35010202), the Special Foundation for Theoretical Physics Research Program of China (Grant No. 11647165), the Fundamental Research Funds for the Central Universities (Grant No. 2020CDJQY-Z003), the National Key R \& D Program of China (Grant No. 2017YFA0304501), the 111 Project (Grant No. D18001), the Hundred Talent Program of the Shanxi Province (2018), and the EMPIR-USOQS, EMPIR Project co-funded by the European Unions Horizon2020 Research and Innovation Programme and the EMPIR Participating States.

\label{ref}

\clearpage
\begin{widetext}
\begin{center}
\begin{large}
\textbf{Supplementary Material of \\``Rabi Spectroscopy and Sensitivity of a Floquet Engineered Optical Lattice Clock"}
\end{large}
\end{center}

\section*{S1. Experimental process.}
\subsection*{S1.1 The two-stage of Doppler cooling}
After being ejected outside an oven heated up to $500$$^{\circ}$C, the $Sr$ atoms gas is collimated by a group of tiny metal tubes. In order to transversely cool and collimate the atomic beam, a two-dimensional (2-D) ($x$-$z$ plane) collimating light with wavelength equal to $461\ nm$ is retro-reflected by the mirror to form the 2-D optical molasses. A $10 \ cm$ long Zeeman slower with $10$ coils and a $461 \ nm$ slowing light along the axis of $x$ with power of $55 \ mW$ are employed for reducing the velocity of the atom beam to $50 \ m/s$. The $ 461 \ nm$ laser produced through a second-harmonic generation from a $922 \ nm$ external cavity diode laser (ECDL) is locked to the ($5s^2$)$^1$$S_0\rightleftarrows$($5s5p$)$^1$$P_1$ transition of the $^{88}Sr$ atomic beam, owing to the largest natural abundance of $82.56\%$, and exhibits a detuning of $51.8$ $MHz$ from the ($5s^2$)$^1$$S_0$ $(F = 9/2)$$\rightleftarrows$($5s5p$) $^1$$P_1$$(F = 11/2)$ transition of $^{87}Sr$. The frequency of Zeeman slowing laser is detuned form the transition of ($5s^2$) $^1$$S_0$$(F = 9/2)$$\rightleftarrows$($5s5p$)$^1$$P_1$ $(F = 11/2)$ by $-560 \ MHz$ to compensate the Doppler frequency shift. After the preliminary cooling process, the atomic beam enters into the main trapping chamber and is further cooled by the two-stage magneto-optical trap (MOT) which employs a pair of anti-Helmholtz coils for generating a zero magnetic field at the MOT center and nonzero one away from the center which changes linearly with distance from the center with an axial magnetic field gradient of $50 \ Gs/cm$. Meanwhile, the three-dimensional (3-D) optical molasses combining with this magnetic confinement are participating in for two-stage Doppler cooling. 

The first cooling stage works on the strong dipole-allowed ($5s^2$)$^1$$S_0$$(F = 9/2)$$\rightleftarrows$($5s5p$) $^1$$P_1$ $(F = 11/2)$ transition with $32 \ MHz$ natural linewidth, and the frequency of the related $461 \ nm$ trapping laser should be detuned $-40 \ MHz$ to compensate the Doppler frequency shift. Furthermore, in order to repump the atoms from the metastable states ($5s5p$)$^3$$P_2$  (five hyperfine states with individual state splitting on the order of GHz) and$ (5s6s) ^3$$S_1$  (three hyperfine states with states separations of $2-3 \ GHz$) to the ground state ($5s^2$)$^1$$S_0$, a $679 \ nm$ and $707 \ nm$ ECDL are applied for constructing the closed transition.
Both repumping lasers are frequency modulated by driving the piezo actuator (PZT) in order to make the range of its frequency cover all hyperfine levels of the repumping spectrum (spanning more than several gigahertz). This process could greatly increase the number of trapped atoms in the blue MOT by a factor of $15$ approximately. The temperature of the atoms in blue MOT is approximately $5 \ mK$ and the number of the blue MOT atoms is around $2.3\times10^7$.

In the second stage of Doppler cooling, because of the hyperfine structure of $^{87}Sr$ system, in order to raise the efficiency of cooling and trapping, the inter-combination transition of ($5s^2$)$^1$$S_0$$(F = 9/2)$$\rightleftarrows$($5s5p$) $^3$$P_1$$(F = 11/2)$ at $689 \ nm$ ($7.5 \ kHz$ natural linewidth) is applied for the trapping laser, but also a  stirring laser which slightly red-detuned from the transition of  ($5s^2$)$^{1}S_{0}$ $(F = 9/2)$ $\rightleftarrows $ $ (5s5p) ^3$$P_{1}$ $(F=9/2)$ at $689 \ nm$ is applied to rapidly randomize the population among the different spin states \cite{Katori_2003}. Both the trapping laser and stirring laser are sourced from a $689 \ nm$ ECDL (the master laser), which is locked to an ultra-low-expansion (ULE) cavity with a fineness of about $10000$ at $689 \ nm$ by using the Pound-Drever-Hall (PDH) technique. The linewidth of the $689 \ nm$ seed laser is approximately $300 \ Hz$, which is suitable for a narrow-line cooling. To reach the required laser power, two slave $689 \ nm$ diode lasers are injection-locked into the master laser. One laser is used for the trapping, the other one is split into two paths, for stirring and pumping, respectively. The pump light at $689 \ nm$ with the $\sigma^+$ polarizations switched by a liquid crystal variable wave-plate are used for pumping atoms from ten Zeeman sublevels to the stretched state of $m_F=+9/2$ before the clock interrogation. After two-stage Doppler cooling, the number of atoms in the red MOT is around $3.5\times10^6$.

\subsection*{S1.2 Loading atoms into the driven optical lattice and Floquet Rabi spectrum probing.}
After two sequential three-dimensional magneto-optical traps cooling, $^{87}Sr$ atoms are cold enough to be loaded in the optical lattice in the Lamb-Dicke regime. Floquet engineering on the clock states is realized by periodically modulating the frequency of the  incident lattice laser ($P_0\approx 300 \ mW$), where a build-in digilock module is used to control the voltage applied on the grating through the PZT. The multi-mode driving is implemented with the help of external function generator. The modulation amplitude $\omega_{m}$ is lowered to a few $GHz$ to guarantee that the piezo actuator works in a linear region. 

The clock transition is interrogated using the $698$ $nm$ clock laser and probed with the $461 \ nm$ laser. A normalized shelving detection method \cite{Katori_2003} is used for enhancing the signal-to-noise ratio and decreasing the influence of the atomic number fluctuation. This method including four steps: (1) exciting the atom from $^1$$S_0$ to excited state $^3$$P_0$ by using clock laser; (2) exciting the atom from $^1$$S_0$ to fluorescence level $^1$$P_1$ by using $461$ $nm$ laser for counting the number $N_1$ of residue atoms at ground state; (3) repumping the atoms on $^3$$P_0$ back to $^1$$S_0$; (4) exciting the atom from $^1$$S_0$ to fluorescence level $^1$$P_1$ again by using $461 \ nm$ laser, and equivalently counting the number $N_2$ of atoms at excited state. The value of spectrum is $N_2/(N_1+N_2)$. The energy levels is shown in Fig.S1 in detail. The propagation direction of the clock laser coincides with the lattice laser by transmissing through the $0$$^{\circ}$ \ dichroic mirror with both HR-coated for $813 \ nm$ and AR-coated for $698 \ nm$. The polarizations of the excitation and lattice beams are oriented in the same direction (for $\pi$-polarization relative to the bias B-field). In order to excite atoms homogeneously, the waist diameter of the $698 \ nm$ excitation laser beam is adjusted to $2000 \ \mu m$ (roughly measurement), $20$ times larger than the focus diameter of the lattice beam.

 The resolved sideband spectrum with typical carrier-sideband structure is obtained by scanning the frequency of the excitation laser through shifting the radio frequency applied to the acousto-optic modulator (AOM) with a step of $200 \ Hz$ in each clock circle. Because the lattice laser operates at the magic wavelength, the transition of carrier peak between the same outer energy levels of the harmonic potential is free of the motional effect. The power of excitation laser at $698 \ nm$ is around $500 \ \mu W$. Due to the saturation broadening, the linewidth of carrier peak is about $1 \ kHz$. The longitudinal trap frequency of harmonic trap potential along the $z$ axis is around $64.8 \ kHz$.

Because of the long lifetime of the excited state $^3$$P_0$, the natural linewidth of the dipole-forbidden transition is around $1 \ mHz$. The absolute frequency of the clock transition is $429228004229873 \ Hz$. To improve the spectral resolution for observing this extremely narrow line-width transition, it is important to stabilize the clock laser at $698 \ nm$ (DL pro from Toptica, Inc.) by locking it to an ultra-low-expansion cavity (made by Stable Laser System) with the help of PDH technique. The fineness of the cavity is near $400,000$ and the length is $10 \ cm$. A fiber phase-noise-cancellation system is employed to suppress the linewidth broadening caused by the delivery fiber. This system is based on a heterodyne Michelson interferometer with the round-trip fiber phase locked onto the phase of a local reference arm. The linewidth of the $698 \ nm$ clock laser is narrowed down to be $1 \ Hz$. The frequency of the clock laser sweeps across the resonant frequency of the transition by using acoustic-optical modulator. For the interrogation time with $150 \ ms$, the Fourier-transform-limited linewidth is $\approx 6 \ Hz$ and the measured linewidth of the spin polarized  stretched state of $m_F=+9/2$ transition is around $6 \ Hz$ measured at the excitation power of $220 \ nW$, which is so close to the Fourier-transform-limit.

\begin{figure*}[h]
	\centering
	\includegraphics[width=0.6\linewidth]{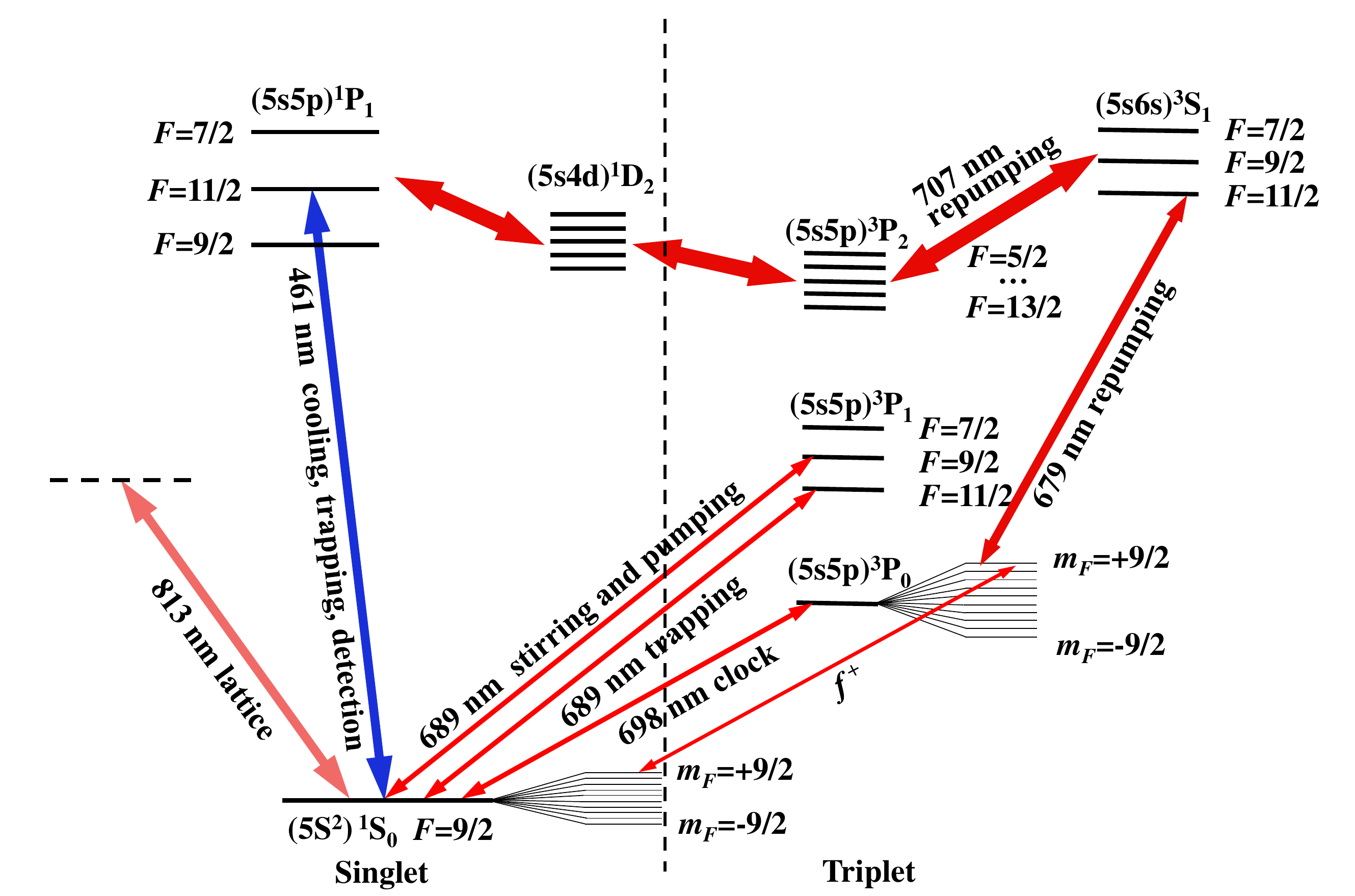}
	\begin{flushleft}
		Figure S1: The schematic of energy levels of the $^{87}Sr$ atoms.
	\end{flushleft}
\end{figure*}

\section*{S2. Theoretical model.}
For the driven system, if we set the position of the retro-reflected mirror as the origin of the coordinate ($z=0$), the light field of incoming and reflecting lattice laser in $z$ direction can be respectively described as 
 \begin{eqnarray}
 &&\vec{E}_{1}=\vec{A}e^{-i\int_{0}^{z}\frac{\omega_{L}(t+z'/c)}{c}dz'+i\int_{0}^{t}\omega_{L}(\tau)d\tau} \\ \nonumber
 &&\vec{E}_{2}=-\vec{A}e^{i\int_{0}^{z}\frac{\omega_{L}(t-z'/c)}{c}dz'+i\int_{0}^{t}\omega_{L}(\tau)d\tau},
 \end{eqnarray} 
 in which $c$ is the speed of light in the vacuum, $\vec{E}_{2}$ picks up a minus sign because of the half wave loss due to reflection. The intensity of the laser is proportional to the square of total field, i.e. $I \propto |\vec{E}_{1}+\vec{E}_{2}|^2/2=2|\vec{A}|^2 \sin 
 ^2\left(\frac{\bar{\omega}_{L}z}{c}+\frac{\omega_1}{\omega_s}\cos(\omega_s t)\sin(\omega_s z/c)\right)$.
 Because $\omega_{s}z/c\sim 10^{-7}$ is very small (we use approximation $\sin(\omega_{s}z/c)\approx\omega_{s}z/c$), the intensity of lattice laser in $z$ direction can be described as 
 \begin{equation}
 I\approxeq I_{0}\sin^{2}\left(\frac{\bar{\omega}_{L}}{c}z+\frac{\omega_{1}}{c}z\sin \omega_{s}t\right).
 \end{equation} 
 We can rewrite $z$ with $L+\Delta z$, in which $L\approx 0.3 \ m$ is the distance between the center of lattice and the high-reflection mirror and $|\Delta z|$ is smaller than the length of the lattice $l=0.4 \ mm$ (means we only consider the region of parameter $z$ in the lattice). After that, because the maximum phase shift caused by $\omega_{1}\Delta z/c$ is of the order of $10^{-3}$, the second term $\frac{\omega_{1}}{c}z\sin \omega_{s}t$ can be replaced with $\frac{\omega_{1}}{c}L\sin \omega_{s}t$. Thus the intensity could be approximately written as
 \begin{equation}
 I\approx I_{0}\sin^{2}\left(\frac{\bar{\omega}_{L}}{c}(z+\frac{\omega_{1}L}{\bar{\omega}_{L}}\sin \omega_{s}t)\right).
 \end{equation} 
 The polarizability of atomic states depends on the frequency.  Since the wavelength of lattice laser is also periodically driven, the differential polarizability of $^{1}S_{0}$ and $^{3}P_{0}$ states is no longer zero. Because $\omega_1/\bar{\omega}_L \ll 1$, we could keep first order truncation around the frequency $\bar{\omega}_L$ corresponding to the magic wave length $\lambda_L$:
 \begin{eqnarray}
 V_{s}&\approx&\left(\alpha_{0}+k_{s}\omega_{1}\sin \omega_{s}t\right)\frac{I}{2\epsilon_{0} c}=\alpha_{s}(t)\frac{I}{2\epsilon_{0} c}, \\ \nonumber
 V_{p}&\approx&\left(\alpha_{0}+k_{p}\omega_{1}\sin \omega_{s}t\right)\frac{I}{2\epsilon_{0} c}=\alpha_{p}(t)\frac{I}{2\epsilon_{0} c},
 \end{eqnarray}
 in which $V_{s}$ and $V_{p}$ are respectively the potential of $^{1}S_{0}$ and $^{3}P_{0}$ states, $k_{s}=\frac{d \alpha_{s}(\omega)}{d\omega}|_{\omega_L}$ and $k_{p}=\frac{d \alpha_{p}(\omega)}{d\omega}|_{\omega_L}$ are respectively the first derivative of polarizability at the magic wavelength, $\epsilon_{0}$ is the permittivity of vacuum and $\alpha_{0}$ is the polarizability at the magic wavelength. Because of Pauli exclusion principle, the interaction between atoms could be ignored. If we only consider the motion in $z$ direction (the lattice direction), the Hamiltonian of the system could be written in the laboratory frame as \begin{equation}
 \hat{H}_{lab}=-\frac{\hat{p}^2}{2m}+\frac{V_{s}+V_{p}}{2}+\frac{V_{p}-V_{s}}{2}\hat{ \sigma}^{(3)}_{\vec{n}}+\frac{\hbar \omega_0}{2}\hat{ \sigma}^{(3)}_{\vec{n}}+g\cos \omega_pt \hat{ \sigma}^{(1)}_{\vec{n}}, \label{hlab}
 \end{equation}
 in which $\hat{\sigma}_{\vec{n}}$ are the Pauli matrix for external state $\vec{n}$, $\hbar \omega_0$ is the energy difference between $^{1}S_{0}$ and $^{3}P_{0}$ states, $g$ is the coupling strength between the clock laser and the atom which depends on the external states, and $\omega_p$ is the angular frequency of the clock laser which assuming aligned perfectly with lattice laser (the effect of misalignment will be discussed later). We first construct a unitary operator $\hat{U}=e^{-i \frac{\omega_0 t}{2}\hat{ \sigma}^{(3)}_{\vec{n}}}$ ($e^{i\theta \hat{ \sigma}^{(3)}_{\vec{n}}\hat{ \sigma}^{(3)}_{\vec{n}}}\hat{ \sigma}^{(1)}_{\vec{n}}e^{-i\theta \hat{ \sigma}^{(3)}_{\vec{n}}}=\hat{ \sigma}^{(1)}_{\vec{n}}\cos(2\theta)+\hat{ \sigma}^{(2)}_{\vec{n}}\sin(2\theta)$ ), $\hat{U}^{\dagger}(\hat{H}_{lab}-i\hbar \frac{\partial}{\partial t})\hat{U}$ gives the Hamiltonian in interaction pictures, ignoring all the counter-rotating wave terms (related to $\omega_0+\omega_p$) and transform back with $\hat{U}^{'}=e^{i\frac{(\omega_0-\omega_p)t}{2}\hat{ \sigma}^{(3)}_{\vec{n}}}$ gives us the rotating wave approximation Hamiltonian 
 \begin{equation}
 \hat{H}_{RWA}=-\frac{\hat{p}^2}{2m}+\frac{V_{p}+V_{s}}{2}+\frac{V_{p}-V_{s}}{2}\hat{ \sigma}^{(3)}_{\vec{n}}+\frac{\hbar \omega_0-\hbar \omega_p}{2}\hat{ \sigma}^{(3)}_{\vec{n}}+\frac{g}{2} \hat{ \sigma}^{(1)}_{\vec{n}}, \label{RWA}
 \end{equation}.
 
 To see Eq.\ref{RWA} more clear, we now move to co-moving frame. To do that, we first construct a unitary operator $\hat{U}_1=exp(\frac{-i\hat{p}}{\hbar}\frac{\omega_{1}L}{\bar{\omega}_{L}}\sin \omega_{s}t)$, which means the corresponding position shift in $z$ direction,  i.e. $\hat{U}_1 z \hat{U}_1^{\dagger}=z-\frac{\omega_{1}L}{\bar{\omega}_{L}}\sin \omega_{s}t$. $\hat{U}_1 (\hat{H}_{RWA}-i\hbar \partial/\partial t) \hat{U}_1^{\dagger}$ gives us the new Hamiltonian after the rotation
 \begin{eqnarray}
 \hat{H}_{1}&=&-\frac{(\hat{p}-m\frac{\omega_{1}\omega_{s}L}{\bar{\omega}_{L}}\cos\omega_{s}t)^2}{2m}+\frac{\alpha_{s}(t)+\alpha_{p}(t)}{4\epsilon_{0}c}I_{0}\sin^{2}\left(\frac{\bar{\omega}_{L}}{c} z\right)+\frac{\hbar (\omega_0-\omega_{p})}{2}\hat{ \sigma}^{(3)}_{\vec{n}} \\\nonumber
 &&+\frac{\alpha_{p}(t)-\alpha_{s}(t)}{4\epsilon_{0}c}I_{0}\sin^{2}\left(\frac{\bar{\omega}_{L}}{c} z\right)\hat{ \sigma}^{(3)}_{\vec{n}}+\frac{g}{2} \hat{ \sigma}^{(1)}_{\vec{n}}+\frac{m}{2}\left(\frac{\omega_{1}\omega_{s}L}{\bar{\omega}_{L}}\cos\omega_{s}t\right)^2.
 \end{eqnarray}
 We could see that the atom pick up additional kinetic energy with the speed $v(t)=-\frac{\omega_{1}\omega_{s}L}{\bar{\omega}_{L}}\cos\omega_{s}t$. Then we construct a unitary operator $\hat{U}_2=\exp(\frac{i}{\hbar}mv(t)\hat{z})$, which means the corresponding momentum shift, i.e.  $\hat{U}_2 \hat{p} \hat{U}_2^{\dagger}=\hat{p}-mv(t)$. One should also notice that the momentum shifts changes the frequency of clock laser due to Doppler effect (special relativity) that  $k_{p}^{'}=(1-v(t)/c) k_{p}$. Thus $\hat{U}_2 (\hat{H}_{1}-i\hbar \partial/\partial t) \hat{U}_2^{\dagger}$ gives us a new Hamiltonian
 \begin{eqnarray}
 \hat{H}_2&=&-\frac{\hat{p}^2}{2m}+\frac{\alpha_{s}(t)+\alpha_{p}(t)}{4\epsilon_{0}c}I_{0}\sin^{2}\left(\frac{\bar{\omega}_{L}}{c} z\right) +\frac{\alpha_{p}(t)-\alpha_{s}(t)}{4\epsilon_{0}c}I_{0}\sin^{2}\left(\frac{\bar{\omega}_{L}}{c} z\right)\hat{ \sigma}^{(3)}_{\vec{n}} \\\nonumber
 && +\frac{\hbar \omega_0}{2}\hat{ \sigma}^{(3)}_{\vec{n}}+\frac{g}{2} \hat{ \sigma}^{(1)}_{\vec{n}}-\frac{\hbar \omega_p(1-\frac{v(t)}{c})}{2}\hat{ \sigma}^{(3)}_{\vec{n}}+\frac{m}{2}v^2(t)-m\frac{\omega_{1}\omega_{s}^{2}Lz}{\bar{\omega}_{L}}\sin\omega_{s}t,
 \end{eqnarray}
 and the kinetic energy related term $mv^2(t)/2$ can be removed after energy shift. 
 
 The energy scales of different terms can be analyzed as following. Basing on previous reference \cite{Katori_2003}\cite{Ovsiannikov_2003}, the magic wavelength polarizability is $\alpha_{0}=4.57\times10^{-39} \ mC^2/N$,  $k_{s}= -1.27\times10^{-53} \ mC^2s/N$ and $k_{p}= -4.57\times10^{-53} \ mC^2s/N$. Because $-(k_a+k_s)\omega_{1}/\alpha_{0}\approx10^{-5}$, the second term  $\frac{\alpha_{s}(t)+\alpha_{p}(t)}{4\epsilon_{0}c}I_{0}$ is approximately $\frac{\alpha_{0}I_0}{2\epsilon_{0}c}=\frac{4\alpha_{0}P_0}{\pi\epsilon_{0}cW^2_0}$ where $P_0$ is the power of the lattice laser, $W_0$ is the waist radius. When $P_0$ is $300 \ mW$ and $W_0$ is $50 \ \mu m$, the second term $(\alpha_{s}(t)+\alpha_{p}(t))I_0/(4\epsilon_{0}c)$ is $276 \ kHz$ in unit of $h$. Meanwhile the third term $(\alpha_{p}(t)-\alpha_{s}(t))I_0/(4\epsilon_{0}c)$ is about only $1 \ Hz$, so that it can be neglected. On the other hand, the last term $m\frac{\omega_{1}\omega_{s}^{2}Lz}{\bar{\omega}_{L}}\sin\omega_{s}t$ means each lattice site will feel different potential energy, but the deviation per site $m\frac{\omega_{1}\omega_{s}^{2}L\lambda_L}{\bar{\omega}_{L}}$ is just about $0.12 \ Hz$. Thus, it can also be omitted. At last, the term $\hbar \omega_{p} v(t)/c$ should be kept because its amplitude $\frac{\hbar \omega_p \omega_1 \omega_s L}{\bar{\omega}_L c}$ varies from zero to several $\hbar \omega_s$. 
 
 The previous discussion focuses on the motion at $z$ direction. In a realistic system, the transverse motion should also be considered since the clock laser might have a small misaligned angle $\delta \theta$ from the lattice laser. Thus our system should be described by a quasi one dimensional optical lattice, in which 'quasi' means that it is the length of lattice in $z$ direction around $0.4 \ mm$, while in $r$ direction it is a Gaussian potential with beam waist diameter of $100 \ \mu m$. With above simplification the system Hamiltonian could be written as $\hat{H}=\hat{H}_{\mathrm{ext}}+\hat{H}_{\mathrm{int}}$, in which
 \begin{eqnarray}
 \hat{H}_{\mathrm{ext}}&=&-\frac{\hat{\vec{p}}^2}{2m}+V_{0}\cos^{2}\left(\frac{\bar{\omega}_{L}}{c} z\right)e^{-2r^2/W_{0}^2},\label{hal} \\\nonumber
 \hat{H}_{\mathrm{int}}&=&\left(\frac{ \hbar\delta}{2}+A \hbar \omega_s \cos \omega_st\right)\hat{ \sigma}^{(3)}_{\vec{n}}+\frac{g}{2}\hat{ \sigma}^{(1)}_{\vec{n}}, \label{Hal}
 \end{eqnarray}
 in which $\delta= \omega_{0}-\omega_{p}$ is the detuning angular frequency, $A=\frac{\omega_p \omega_1  L}{2\bar{\omega}_L c}$ is the renormalized driven amplitude, which is irrelevant with $\omega_s$, $V_0=\frac{4\alpha_{0}P_0}{\pi\epsilon_{0}cW^2_0}=h\nu_0$ ($\nu_0=276 \ kHz$) is the lattice depth, and $r=\sqrt{x^2+y^2}$ designates the transverse distance from the lattice axis.
 
 In our system, the lattice depth $V_0$ is about $90 \ E_{R}$ , in which $E_{R}=\frac{h^2}{2m\lambda_L^2}=h\nu_{rec}$ ($\nu_{rec}=3.44 \ kHz$) is the recoil energy, thus we could ignore the hopping between different lattice sites. Because of the angular momentum conservation law, we only need to consider the motion in the same planer with the clock laser. With quadratic approximation of Eq.\ref{hal}, the external Hamiltonian could be written as 
 \begin{equation}
 \hat{H}_{\mathrm{ext}}=-\frac{\hat{\vec{p}}^2}{2m}+V_{0}\left(\frac{\bar{\omega}_{L}^2}{c^2} z^2 +\frac{2}{W_{0}^2}r^2\right).
 \end{equation}
 The corresponding eigen function is $|n_z,n_r\rangle$, and the eigen energy is 
 \begin{equation}
 E_{n_{z},n_{r}}=h\nu_{z}(n_z+1/2)+h\nu_{r}(n_r+1),
 \end{equation}
 with $\nu_{z}=\sqrt{\frac{V_0 \bar{\omega}_L^2}{2m\pi^2 c^2}}$, $\nu_r=\sqrt{\frac{V_0}{m \pi^2 W_0^2}}$, and $\nu_{z}/\nu_{r}=\frac{\bar{\omega}_LW_{0}}{c\sqrt{2}}\approx328$.

Because the the transverse trap is isotropic, we choose a small misaligned angle $\delta \theta$ along the propagate direction such that $\vec{k}_p=k_p (\vec{z}+\delta \theta \vec{x})$. Thus the single atom internal Hamiltonian could be written as 
 \begin{equation}
 \hat{H}_{\mathrm{int}}^{n_z,n_x}=(\frac{ \hbar\delta}2+A \hbar \omega_s \cos \omega_st)\hat{ \sigma}^{(3)}_{\vec{n}}+\frac{g_{n_z,n_x}}{2} \hat{ \sigma}^{(1)}_{\vec{n}},
 \end{equation}
 in which $g_{n_z,n_x}=g_0 \langle n_z,n_x | e^{i (k_p z+\delta \theta k_p x)}|n_z,n_x \rangle$ is the coupling strength between atoms and the clock laser in the external state $|n_z,n_x \rangle$.  The further calculation gives:
 \begin{equation}
 \langle n_z,n_r | e^{i (k_p z+\delta \theta k_p x)}|n_z,n_r \rangle=e^{-(\eta_z^2+\eta_x^2)/2}L_{n_x}(\eta_x^2)L_{n_z}(\eta_z^2),
 \end{equation}
 in which $L_{n}$ is the $nth$ order Laguerre polynomial $\eta_z=\sqrt{h/(2m\nu_z)}/\lambda_p $ and  $\eta_x= \sqrt{h/(2m\nu_r)}\delta \theta/\lambda_p$ are the corresponding Lamb-Dick parameters, $\lambda_p \approx 698 \ nm$ is the wave length of the clock laser. This misaligned angle $\delta\theta$ will also affect the re-normalized driven amplitude $A$ as the clock laser vector in $z$ direction is not $k_p$ but $k_p\cos(\delta\theta)$. However, because $\delta\theta$ is smaller than $10 \ mrad$ in our experiment, its effect on $A$ could be ignore.

\section*{S3. Spectroscopy analysis.}
The Hamiltonian $\hat{H}_{\mathrm{int}}$ in Eq.\ref{Hal} could be written in the rotating frame with an unitary transformation 
\begin{equation}
\hat{U}=\exp\left[-i\left(\frac{\delta t}{2}+A\sin \omega_s t\right)\hat{ \sigma}^{(3)}_{\vec{n}}\right].
\end{equation}
Then, the rotating Hamiltonian $\hat{H}_{r}=\hat{U}^{\dagger}(\hat{H}_{\mathrm{int}}-i\hbar \partial/\partial t)\hat{U}$ gives 
\begin{equation}
\hat{H}_r=\frac{g_{\vec{n}}}{2} \left(\exp\left[i\delta t\right]\sum_{k}J_{k}[2A]\exp[ik\omega_{s}t]\hat{ \sigma}^+_{\vec{n}} + h.c.\right),
\end{equation}
in which $J_{k}[\hspace{3px}]$ is the $k$th order first kind of Bessel function. If the driven frequency $\omega_s$ is much larger than the Rabi frequency $g_0/\hbar$, all Floquet side bands can be treated as been decoupled which is named resolved Floquet side band approximation (RFSBA). After that, We could transform back with the unitary operator $\hat{U}'=\exp[\frac{it}{2}\left(\delta-k \omega_s \right)\hat{ \sigma}^{(3)}_{\vec{n}}]$ and obtain the effective Hamiltonian of the $k$th order Floquet side band
\begin{equation}
\hat{H}_{eff}^{k}=\frac{\hbar\delta-k\hbar\omega_s}{2}\hat{ \sigma}^{(3)}_{\vec{n}}+\frac{g_{\vec{n}}}{2} J_{k}[2A] \hat{ \sigma}^{(1)}_{\vec{n}}.\label{heff}
\end{equation}
If we assuming all the atoms are prepared at ground state $^1$$S_0$, then after diagonalizing above Hamiltonian, we could get the occupation probability of the $^3 P_{0}$ state 
\begin{equation}
P_{e}^{\vec{n}}=\sum_{k}\left(\frac{g_{\vec{n}}J_{k}[2A]}{\hbar R_{k}}\right)^2 \sin^{2}\left[\frac{t}{2}R_{k}\right], \label{pe}
\end{equation}
in which $R_{k}=\sqrt{(g_{\vec{n}}J_{k}[2A]/\hbar)^2+(\delta-k\omega)^2}$. 

Now, we consider the effect of distribution of atoms in the external states. Because the temperature of the system is typically several $\mu K$, we could approximately use normalized Boltzmann distribution to describe the system, thus 
\begin{equation}
P_{e}(t)=\sum_{n_z,n_r}q_{r}(n_{r})q_{r}(n_{r})P_{e}^{n_z,n_r}(t), \label{ptotal}
\end{equation}  
in which 
\begin{eqnarray}
q_{r}(n_{r})&=&\frac{e^{-(n_r+1/2)h\nu_r/(k_B T_r)}}{\sum_{n_r=0}^{N_r-1}e^{-(n_r+1/2)h\nu_r/(k_B T_r)}}, \\\nonumber
q_{z}(n_{z})&=&\frac{e^{-(n_z+1/2)h\nu_z/(k_B T_z)}}{\sum_{n_z=0}^{N_z-1}e^{-(n_z+1/2)h\nu_z/(k_B T_z)}},
\end{eqnarray}
and $N_{r(z)}=\frac{V_0}{h\nu_{r(z)}}$ are the number of energy levels of different direction in the trap, $k_B$ is the Boltzmann constant. The number of motional state chosen in the calculation is determined by the depth of traps.

After substituting Eqn.\ref{pe} into Eqn.\ref{ptotal}, we obtain the explicit form of the excited state probability under RFSBA:
\begin{eqnarray}
P_{e}^{k}(\delta,t)=\sum_{n_z=0}^{N_z-1}\sum_{n_r=0}^{N_r-1}q_{r}(n_{r})q_{z}(n_{z})\sum_{k}\left(\frac{\sin\left[t R_{k}/2\right]\Gamma_{k}}{R_{k}}\right)^2, \label{explit}
\end{eqnarray}
in which 
\begin{eqnarray}
&&q_{r}(n_{r})=\frac{e^{-(n_r+1/2)h\nu_r/(k_B T_r)}}{\sum_{n_r=0}^{N_r-1}e^{-(n_r+1/2)h\nu_r/(k_B T_r)}}, \\\nonumber
&&q_{z}(n_{z})=\frac{e^{-(n_z+1/2)h\nu_z/(k_B T_z)}}{\sum_{n_z=0}^{N_z-1}e^{-(n_z+1/2)h\nu_z/(k_B T_z)}}, \\\nonumber
&&\Gamma_{k}=g_0e^{-(\eta_z^2+\eta_r^2)/2}L_{n_r}(\eta_r^2)L_{n_z}(\eta_z^2)J_{k}[2A]/\hbar,\\\nonumber
&&R_{k}=\sqrt{\Gamma_{k}^2+(\delta-k\omega_s)^2}, \\\nonumber
&&\eta_z=\sqrt{h/(2m\nu_z)}/\lambda_p, \eta_r= \sqrt{h/(2m\nu_r)}\delta \theta/\lambda_p \text{  and  } A=\kappa \bar{V}.
\end{eqnarray}
Indeed, all the coefficients can be determined in the experiment which will be explained in detail in next Subsection, and those related to our platform are list in the Table.\ref{coe} below.
\begin{table}[h]\begin{center}
		\begin{tabular}{|c|l|c|c|c|c|c|c|}
			\hline
			Parameter & Value & Parameter & Value & Parameter & Value & Constant & Value\\
			\hline
			$T_z$ &  $3.0 \ \mu K$ & 
			$T_r$ &  $3.7 \ \mu K$ & 
			$\nu_z$ &  $65 \ kHz$ &
			m & $87.62 \ u$ \\
			\hline
			$\nu_r$ &  $250 \ Hz$ &
			$N_z$ &  $5$ &
			$N_r$ & $1296$ &
			$\lambda_p$ & $698 \ nm$ \\
			\hline
			$\delta \theta$ &  $9 \ mrad$ &
			$g_0$ &  $h\times 3.3 \ Hz$ &
			$\kappa$ &  $0.76/V$  & &\\
			\hline
		\end{tabular}
		\caption{The determined system parameters and some constants.}\label{coe}
\end{center}\end{table}

With coefficient Table.\ref{coe} and explicit formula Eqn.\ref{explit}, we can calculate Floquet Rabi spectrum of different frequency modulation coefficients including the FM frequency $\omega_s$ and the mean value of the voltage $\bar{V}$.

\section*{S4. Extraction of the experimental parameters.}
To get the time evolution of the total probability of the $^{3}P_0$ state, one needs to know the system parameters $\nu_z$, $\nu_r$, $N_z$, $N_r$, $T_z$, $T_r$, $\delta \theta$, $g_0$. Because the driven frequency $\omega_s \ll 2\pi \nu_z$, we assume that it will not change the atom distribution on the external states.  Also the Time-of-Flight (TOF) pictures in our experiment shows that no obvious atom loss due to driven, thus we could determine these parameters by detecting the Rabi spectrum without driven. 

By doing this, we could first use a large power clock laser in $z$ direction with saturation broadening to get the motional side band spectrum in $z$ direction.  We could see from Fig. S2 (a) that both blue and red side band spectrum have complicated shapes, thus indicates that second order approximation of $\hat{H}_{\mathrm{ext}}$ is not enough. The fourth order approximation gives us the external energy level:
\begin{equation}
E_{n_r,n_z}/h=\nu_z \left(n_z+1/2\right)+\nu_r \left(n_r+1\right)-\frac{\nu_{rec}}{2}\left(n_z^2+n_z+1/2\right)-\nu_{rec}\frac{\nu_r}{\nu_z}\left(n_r+1\right)\left(n_z+1/2\right).
\end{equation}
thus the maximum energy gap in $z$ direction is $\Delta_{z}=(E_{0,1}-E_{0,0})/h=\nu_z-\nu_{rec}$ and the lattice recoil frequency $\nu_{rec}=\frac{h}{2m\lambda_L^2}=3.44 \ kHz$. We could read from Fig. S2 $(a)$ that $\nu_z \approx 65 \ kHz$, which matches well with quadratic approximation value $\sqrt{\frac{V_0 \bar{\omega}_L^2}{2m\pi^2 c^2}}=61.6 \ kHz$. The number of energy level $N_z=\frac{\nu_0}{\nu_z}\approx 5$. The red side band $(n_z\rightarrow n_z-1)$ is suppressed with respect to the blue side band $(n_z\rightarrow n_z+1)$. If one ignore the details of the line shapes of the side band, the ratio of the integrated side band absorption cross section
\begin{equation}
\frac{\sigma_{red}}{\sigma_{blue}}=\frac{\sum_{n_z=1}^{N_z}q_{z}(n_{z})}{\sum_{n_z=0}^{N_z-1}q_{z}(n_{z})}
\end{equation}
gives us the temperature $T_z=3.0 \ \mu K$. 

\begin{figure*}
	\centering
	\includegraphics[width=0.48\linewidth ]{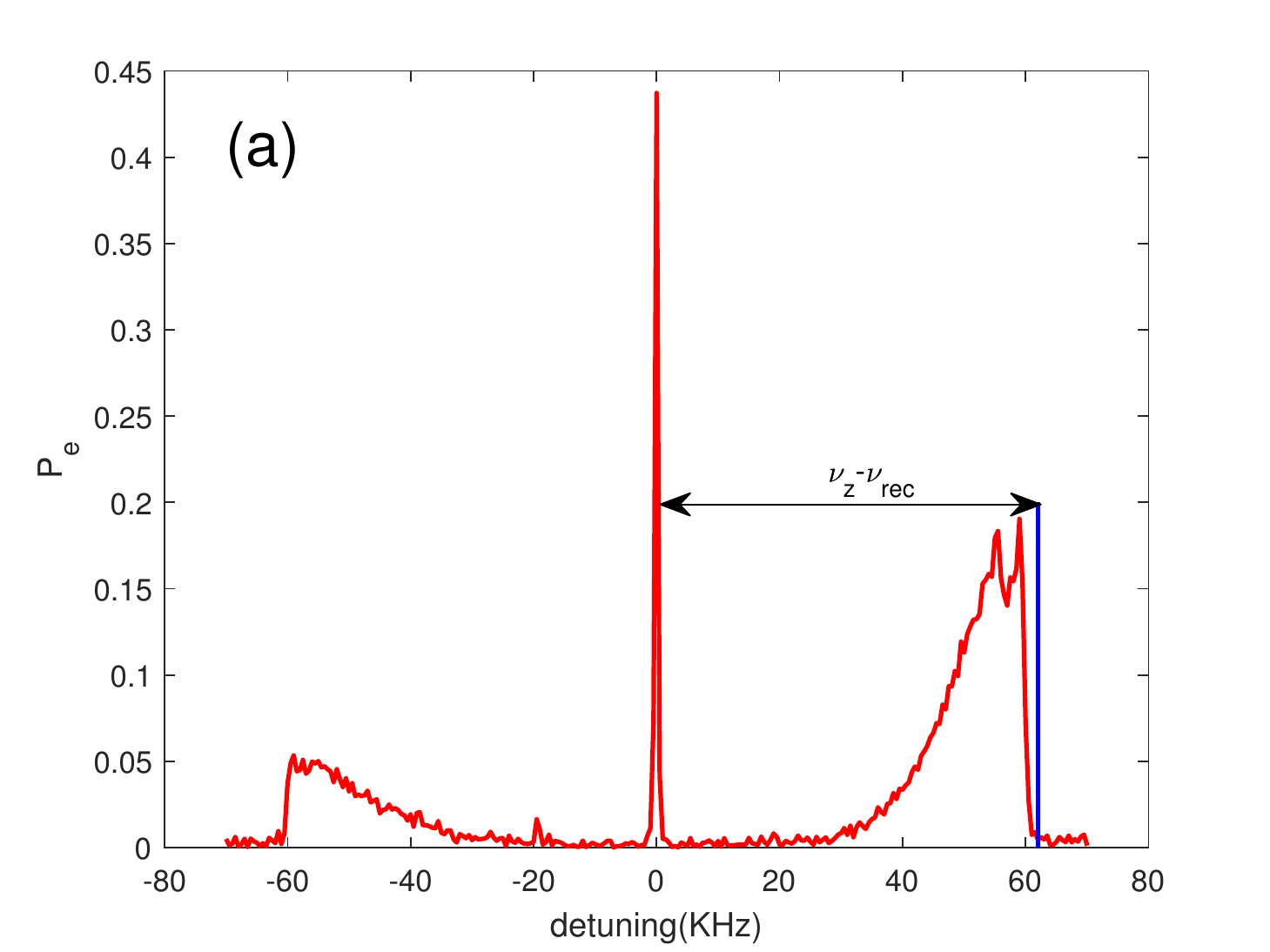}\hfill
	\includegraphics[width=0.48\linewidth]{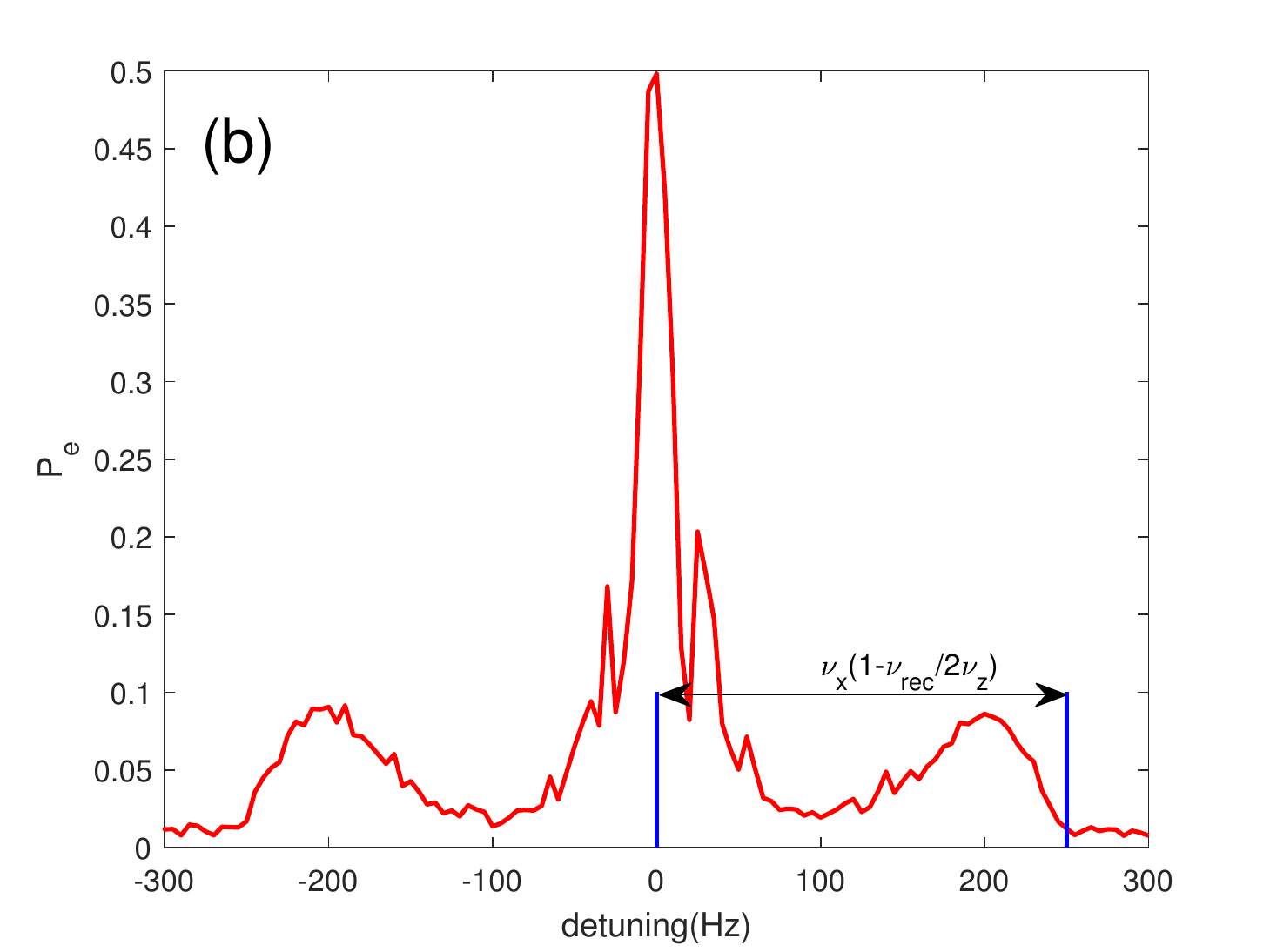} \\
	\includegraphics[width=0.48\linewidth ]{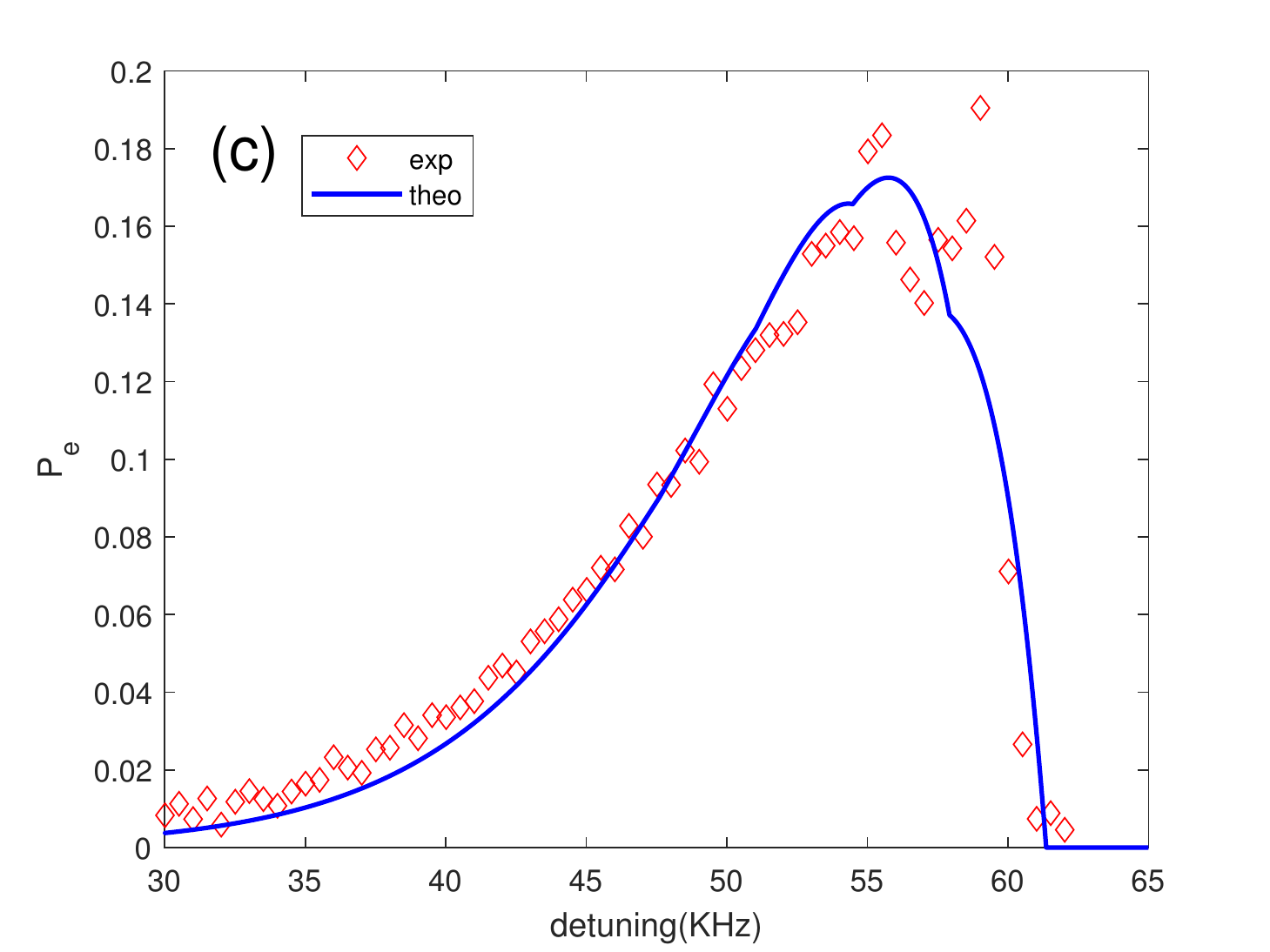}\hfill
	\includegraphics[width=0.48\linewidth]{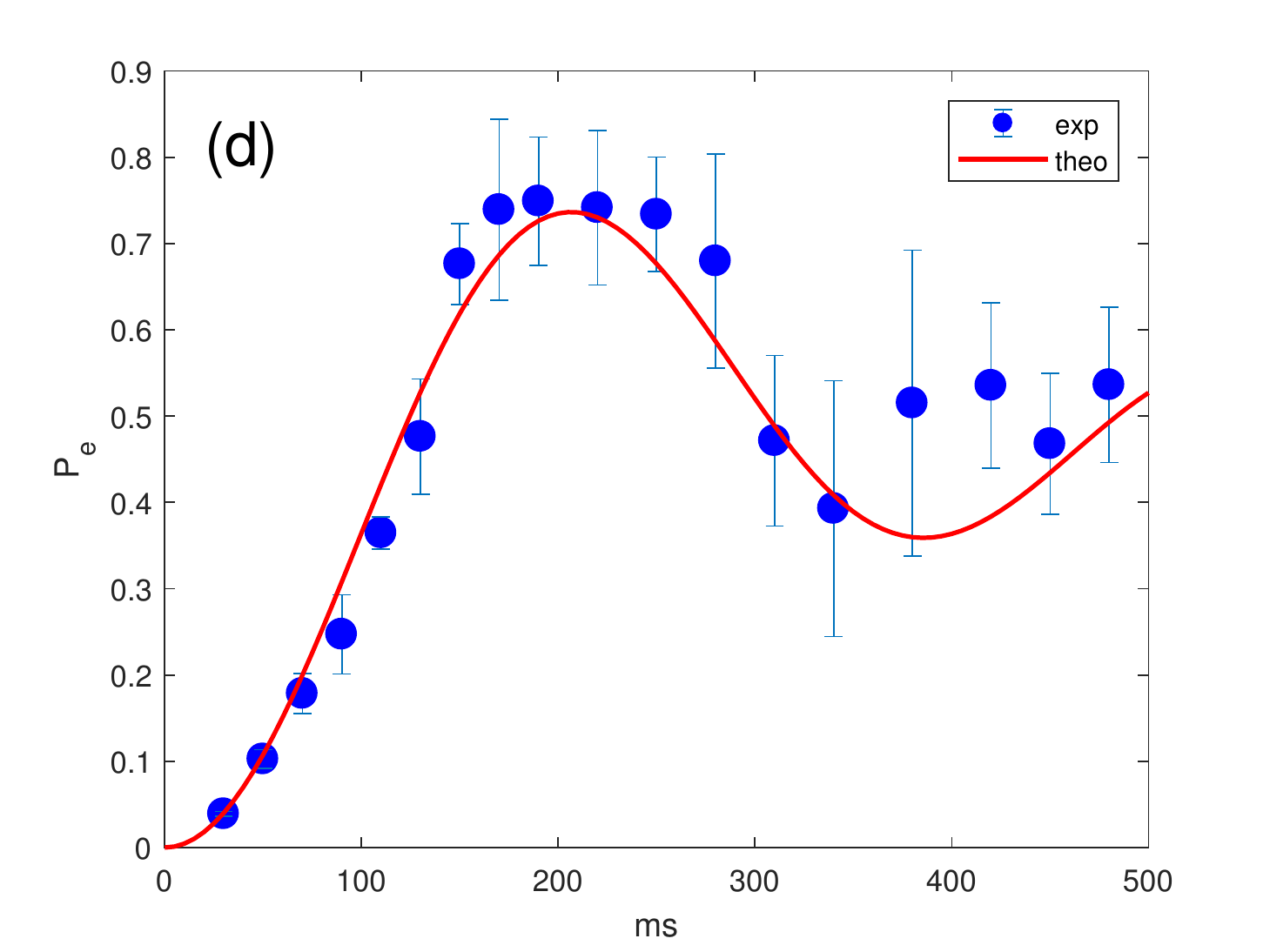}
	\begin{flushleft}
		Figure S2: Determination of the system parameters in undriven case. (a) The side spectrum in $z$ direction. (b) The side spectrum in $x$ direction; (c) The comparison between theoretic fitting and experimental data for blue side band in $z$ direction and (d) the Rabi oscillation. 
	\end{flushleft}
\end{figure*}
Then we could give the clock laser a small misaligned angle (usually several degrees) from $z$ axis to get the side band spectrum in $x$ direction. The maximum energy gap $\Delta_{x}=(E_{0,1}-E_{0,0})/h=\nu_r \left(1-\frac{\nu_{rec}}{2\nu_z}\right)$,
thus we could read from Fig.S2 (b) that $\nu_r \approx 250 \ Hz$, which is not too far away from the quadratic approximation $\sqrt{\frac{V_0}{m \pi^2 W_0^2}}\approx188 \ Hz$. The number of energy levels are $N_r \approx N_z \frac{\nu_z}{\nu_r} \approx 1296$. In principle, one could also get the temperature $T_r$ from the ratio of the side band cross section in $x$ direction. But since $\nu_r$ is small,  the error bar of $T_r$ estimated in this way is really large. Instead, we get $T_r$ from fitting the shape of the blue side spectrum in $z$ direction using equation \cite{Ye_2009_s1}:
\begin{equation}
\sigma_{blue}(\delta\theta) \propto \sum_{n_z=0}^{N_z}q_{z}(n_{z})\left(1-\frac{\delta\theta}{\gamma(n_z)}\right)e^{-\alpha\left(1-\frac{\delta\theta}{\gamma(n_z)}\right)}\Theta\left[\gamma(n_z)-\delta\theta\right]
\end{equation}
with $\gamma(n_z)=\nu_z-\nu_{rec}(n_z+1)$, $\alpha=\left[\gamma(n_z)/\nu_{rec}\right](h\nu_z/k_B T_r)$ and $\Theta$ the Heaviside function. We can read from Fig.S2 (c) that $T_r=3.7 \ \mu K$. 

At last, we determine the parameter $g_0$ and $\delta\theta$ by fitting the rabi-oscillation as shown in Fig.S2 (d). The Rabi oscillation of the non-driven system described by the equation:
\begin{equation}
P_e(t)=\sum_{n_r,n_z}q_{r}(n_{r})q_{z}(n_{z})\sin^2 \left(e^{-(\eta_z^2+\eta_r^2)/2}L_{n_r}(\eta_r^2)L_{n_z}(\eta_z^2)\frac{2g_0 t}{\hbar}\right),
\end{equation}
which gives us that $\delta\theta = 9 \ mrad$ and $g_0/h=3.3 \ Hz$. The damping of the Rabi oscillations is a dephasing caused by finite temperature effects, since atoms at different motional states have different Rabi frequencies. 

All the above parameters could be determined by the undriven system, the only left unfixed parameter is the renormalized driven amplitude $A$, which is proportional to the voltage $\bar{V}$ that added on the PZT, i.e. $A=\kappa \bar{V}$. The coefficient $\kappa$ could be determined also by simulating the Rabi oscillation of any Floquet band of the driven system because all the other parameters are fixed. In Fig.S3, we can find the renormalized amplitude $A$ is perfectly linearly related to $\bar{V}$. Thus, in our experiment, we obtain the coefficient $\kappa=0.76/V$ after linear fitting.

\begin{figure*}[h]
	\centering
	\includegraphics[width=0.5\linewidth]{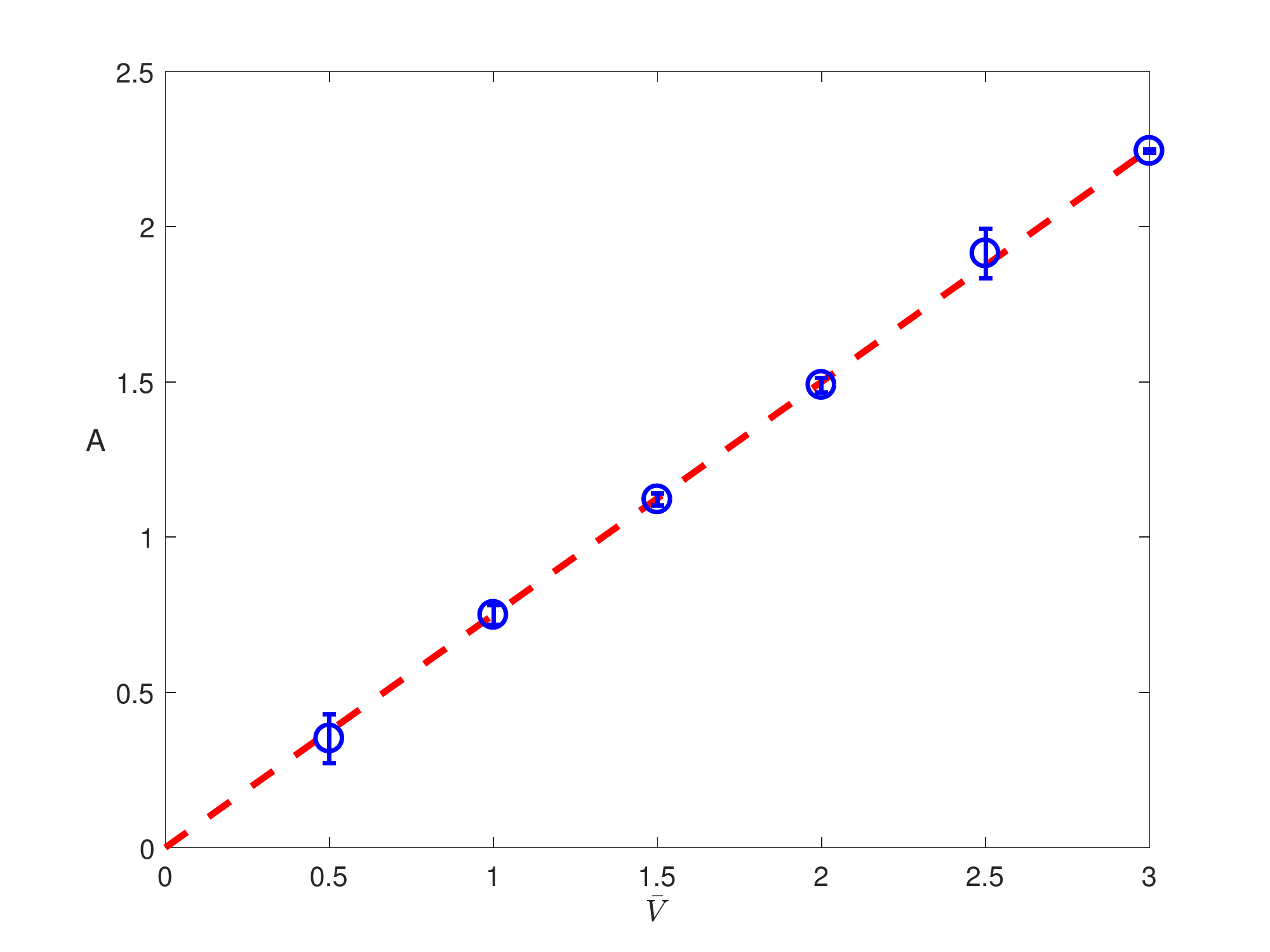}
	
	\begin{flushleft}
		Figure S3: The linear relation between renormalized amplitude $A$ and voltage $\bar{V}$ added on the PZT. 
	\end{flushleft}
\end{figure*}

\section*{S5. The spectrum at different $L$}
From the calculation in section S2., we know that the relation between the renormalized driving amplitude $A$ and the distance $L$ from the reflect mirror to the center of lattice is linear $A=\frac{\omega_p \omega_1  L}{2\bar{\omega}_L c}$. In order to test our theoretical prediction, we implement the experiment of Rabi spectra at different $L$ and voltage $V$ added to the PZT. As shown in the Fig.S4, we set the distance $L$ to be $20 \ cm$, $25.5 \ cm$ and $31.5 \ cm$ and measuring the Rabi spectrum. We found that even with the same voltage added to the PZT, one gets the qualitatively different spectra for different $L$. Demonstrated in the last subfigure (j), we can fit the value of $A$ from the spectra and found that $A$ has a linear dependence of $L$, which is in good agreement with our theoretical prediction.

\begin{figure*}[h]
	\centering
	\includegraphics[width=1\linewidth]{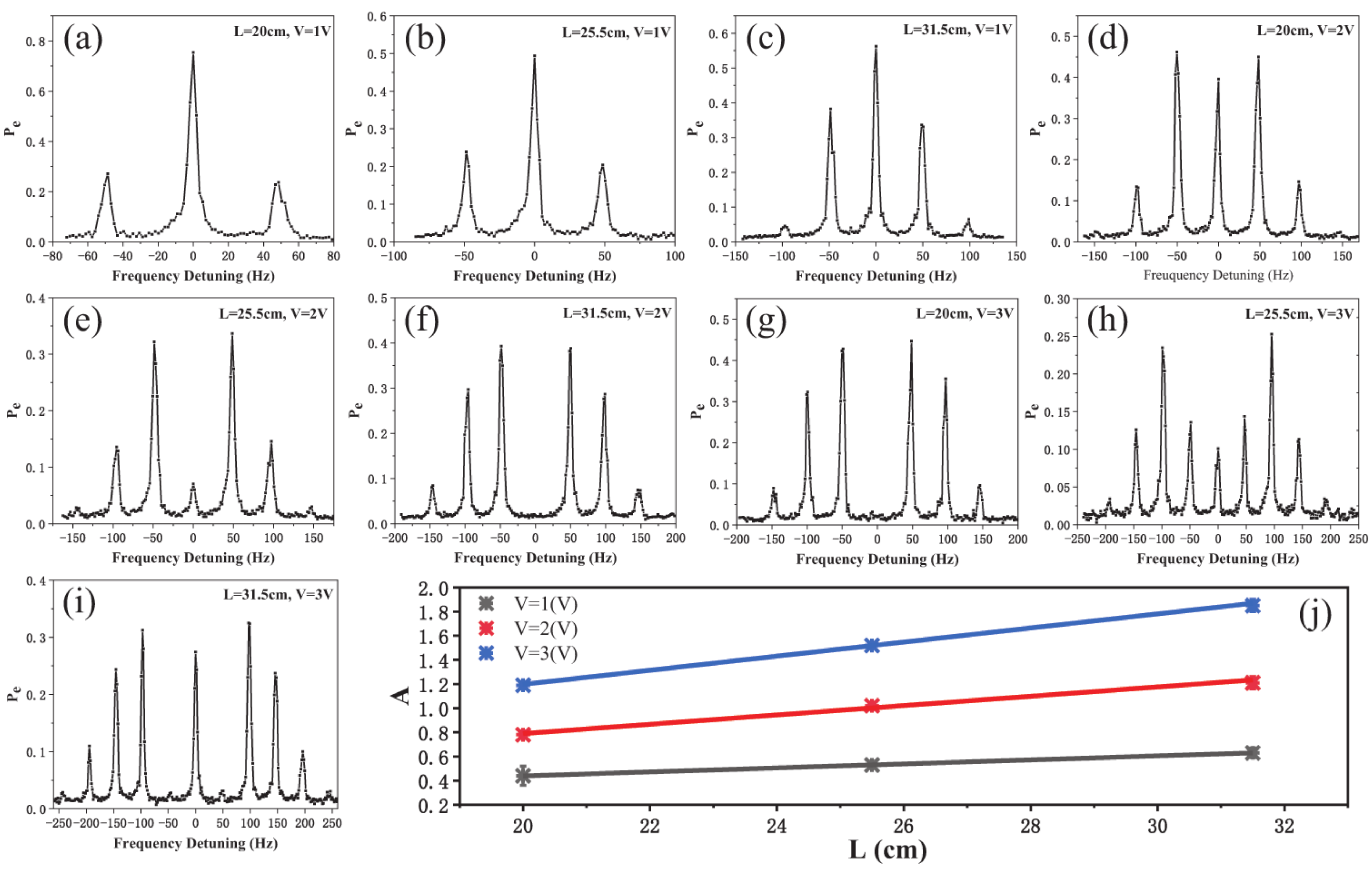}\hfill
	 	\begin{flushleft}
	Figure S4:The Rabi spectra at different voltages $V$ and distance $L$, and the last subfigure (j) presents the relation among driving amplitude $A$, distance $L$ and voltages $V$ adding on the PZT. 
	\end{flushleft}
\end{figure*}

\section*{S6. Fisher information at different Rabi frequencies. }
To extract the Fisher information we fit the experimental data using the trial functions
\begin{equation}\label{sinc2}
f_{\mathrm{sinc^2}}(x)= y_0+A \frac{ \sin^2(\pi t \sqrt{(x-x_c)^2+w^2})}{ (x-x_c)^2+w^2}
\end{equation}
and we therefore calculate the FI with Eq.(7). We report in the figure a typical example of the fit.

 \begin{figure*}[h]
 	\centering
 	\includegraphics[width=0.6\linewidth]{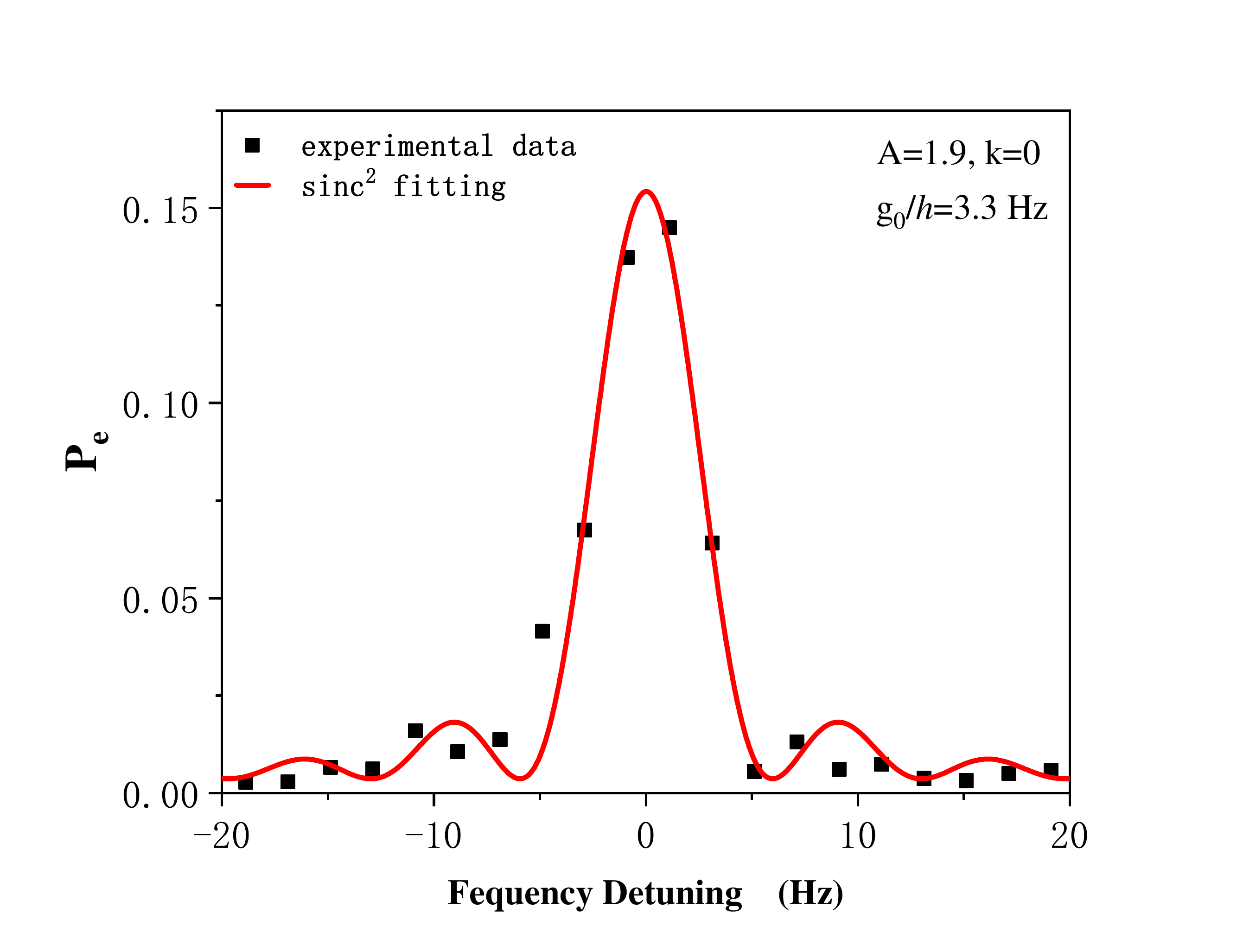}\hfill
 	\begin{flushleft}
 	Figure S5: Experimental Rabi spectrum and the fitting function at $g_0/h=3.3 \ Hz$ for renormalized driven amplitude $A=1.9$ and carrier peak $k=0$.
	\end{flushleft}
 \end{figure*}

\end{widetext}
\end{document}